\documentclass{nsr}

\usepackage{amsmath,graphicx,array,bm,gensymb}
\usepackage{dcolumn,soul}%

\usepackage{amsthm}
\usepackage[figuresright]{rotating}%
\usepackage{algorithm, algorithmicx, algpseudocode}
\usepackage{listings}%
\usepackage{hyperref}

\makeatletter
\def\uns{\ifmmode\,\else$\,$\fi}%

\makeatother
 
\jvol{XX}
\jnum{X}
\jyear{Year}
\doi{10.1093/nsr/XXXX}
\received{XX XX Year}
\revised{XX XX Year}
\accepted{XX XX Year}

\begin{document}

\dhead{REVIEW}

\subhead{PHYSICS}

\title{Quantum geometry in condensed matter}

\author{Tianyu Liu$^{1,2,\dagger}$}

\author{Xiao-Bin Qiang$^{1,2,3,4,\dagger}$}

\author{Hai-Zhou Lu$^{1,2,3,4,*}$}

\author{X. C. Xie$^{5,6,7}$}

\affil{$^1$Department of Physics and Shenzhen Institute for Quantum Science and Engineering, Southern University of Science and Technology (SUSTech), Shenzhen 518055, China}
\affil{$^2$International Quantum Academy, Shenzhen 518048, China}
\affil{$^3$Quantum Science Center of Guangdong-Hong Kong-Macao Greater Bay Area (Guangdong), Shenzhen 518045, China}
\affil{$^4$Shenzhen Key Laboratory of Quantum Science and Engineering, Shenzhen 518055, China}
\affil{$^5$Interdisciplinary Center for Theoretical Physics and Information
Sciences (ICTPIS), Fudan University, Shanghai 200433, China}
\affil{$^6$ International Center for Quantum Materials, School of Physics, Peking University, Beijing100871, China}
\affil{$^7$Hefei National Laboratory, Hefei 230088, China}

\authornote{\textbf{Corresponding author.} Email: luhz@sustech.edu.cn
\\
T.L. and X.-B.Q. contribute equally to this work.}

\abstract[ABSTRACT]{One of the most celebrated accomplishments of modern physics is the description of fundamental principles of nature in the language of geometry. As the motion of celestial bodies is governed by the geometry of spacetime, the motion of electrons in condensed matter can be characterized by the geometry of the Hilbert space of their wave functions. Such quantum geometry, comprising of Berry curvature and quantum metric, can thus exert profound influences on various properties of materials. The dipoles of both Berry curvature and quantum metric produce nonlinear transport. The quantum metric plays an important role in flat-band superconductors by enhancing the transition temperature. The uniformly distributed \textcolor{black}{momentum-space} quantum geometry stabilizes the fractional Chern insulators and results in the fractional quantum anomalous Hall effect. We here review in detail quantum geometry in condensed matter, paying close attention to its effects on nonlinear transport, superconductivity, and topological properties. Possible future research directions in this field are also envisaged.
}



\maketitle

\section{INTRODUCTION}\label{sec1}
\textcolor{black}{Berry curvature \cite{berry1984} has undoubtedly reshaped the modern condensed matter physics \cite{xiaodi2010}.} One prominent example is the family of quantum Hall effects, which have won Nobel prizes in the years 1985, 1998, and 2016. The integer quantum Hall effect \cite{klitzing1980,thouless1982} is the first instance of phase transition beyond the Landau paradigm and marks the commencement of the topological phases of matter, while its fractional \cite{tsui1982, laughlin1983} and anomalous \cite{haldane1988} counterparts are promisingly applicable in quantum computation and dissipationless quantum devices.

Wave packet dynamics \cite{gaoyang2014} has revealed that the polarizability of Berry connection (whose curl is the Berry curvature) also serves as a characteristic geometric quantity and is closely related to the quantum metric \cite{resta2011, torma2023}. Remarkably, quantum metric exerts profound impacts on quantum matter in a way quite similar to that of Berry curvature. In fact, the Berry curvature and quantum metric respectively correspond to the imaginary and real parts of the \textcolor{black}{so-called quantum geometric tensor \cite{resta2011, torma2023, provost1980}, which characterizes the geometry of the Hilbert space comprising of the electron wave functions.}

In this review, we will summarize the role of such a quantum geometric tensor in nonlinear transport, superconductivity, and topology of condensed matter.

Our first focus will be the nonlinear transport \cite{gaoyang2014, sodemann2015, duzz2018, duzz2019, maqiong2019, kangkaifei2019, huangmeizhen2023, tiwari2021, mateng2022, shvetsov2019, kumar2021, dzsaber2021, kaplan2024, wangchong2021, liuhuiying2021, gaoanyuan2023, wangnaizhou2023, wanglujunyu2024}, where an AC input can produce either Hall or longitudinal responses with doubled frequencies. \textcolor{black}{For example}, an AC current in an inversion ($\mathcal P$) broken but time-reversal ($\mathcal T$) symmetric system can give rise to a double-frequency Hall voltage, which is proportional to the dipole of Berry curvature \cite{sodemann2015, duzz2018, duzz2019}. This purely electric nonlinear Hall effect is later realized in a variety of transition metal dichalcogenides (e.g., WTe$_2$ \cite{maqiong2019, kangkaifei2019}, WSe$_2$ \cite{huangmeizhen2023}, and MoTe$_2$ \cite{tiwari2021, mateng2022}) and semimetals (e.g., Cd$_3$As$_2$ \cite{shvetsov2019}, TaIrTe$_4$ \cite{kumar2021}, and Ce$_3$Bi$_4$Pd$_3$ \cite{dzsaber2021}). Remarkably, similar electric nonlinear Hall effect can survive even in the absence of Berry curvature dipoles \cite{gaoyang2014, kaplan2024, wangchong2021, liuhuiying2021, gaoanyuan2023, wangnaizhou2023}. In antiferromagnets that break both $\mathcal P$ and $\mathcal T$ but preserve $\mathcal P\mathcal T$, the Berry curvature dipole is prohibited but the quantum geometry can survive in the form of quantum metric \cite{gaoyang2014, kaplan2024, wangchong2021, liuhuiying2021, gaoanyuan2023, wangnaizhou2023}. The quantum-metric-dipole-induced electric nonlinear Hall effect has been proposed in antiferromagnets CuMnAs \cite{wangchong2021} and Mn$_2$Au \cite{liuhuiying2021}, and has been recently observed in antiferromagnetic topological insulator MnBi$_2$Te$_4$ \cite{gaoanyuan2023, wangnaizhou2023}. Besides the electric nonlinear Hall effect, quantum geometry also plays an important role in the magento-nonlinear Hall effect and nonlinear longitudinal transport, which are respectively realized in kagome magnet Fe$_3$Sn$_2$ \cite{wanglujunyu2024} and MnBi$_2$Te$_4$ \cite{wangnaizhou2023}. All these nonlinear features could serve as new probes detecting the spectral, symmetrical, and topological properties of quantum matter.

We then examine the flat-band superconductors, whose transition temperatures are believed to be dominated by the quantum metric. In particular, for two-dimensional flat-band superconductors, the superfluid weight, and, in turn, the Berezinskii-Kosterlitz-Thouless transition temperature have two sources of contributions: the conventional one vanishing for the dispersionless bands, and the interband-process-induced geometric one which can even survive the perfect flat bands \cite{peotta2015, julku2016, lianglong2017, torma2018, huhtinen2022}. The advert of magic-angle twisted bilayer graphene makes it possible to study the geometric contribution to the superfluid weight and the transition temperature \cite{xiefang2020, huxiang2019, julku2020, tianhaidong2023}. On the theoretical side, the geometric contribution is found to be a Brillouin zone integral of quantum metric. At the mean-field level with isotropic pairing, the superfluid weight of magic-angle twisted bilayer graphene is also isotropic and is bounded by the band topology from below \cite{xiefang2020}. The superfluid weight becomes anisotropic in the presence of strong nearest-neighbor pairing \cite{julku2020}, and its evolution with band filling exhibits dependence on the twist angle \cite{huxiang2019}. On the experimental side, the conventional contribution to the superfluid weight of magic-angle twisted bilayer graphene provides an estimate of the transition temperature \cite{xiefang2020} on the order of $0.1\,\text{K}$, which is much lower than the measured $T_c\simeq2.2\,\text K$ and thus implies a dominating quantum metric contribution \cite{tianhaidong2023}.

We lastly look into the fractional Chern insulators \cite{tang2011, sunkai2011, neupert2011, parameswaran2012, roy2014, jackson2015, wilhelm2021, ledwith2020, repellin2020, abouelkomsan2020, xieyonglong2021, spanton2018, luzhengguang2024, caijiaqi2023, zengyihang2023, park2023, xufan2023}, whose stability is found to subtly rely on the distribution homogeneity of both the Berry curvature and quantum metric \textcolor{black}{in the Brillouin zone}. Following the development from the integer quantum Hall effect \cite{klitzing1980, thouless1982} to the fractional quantum Hall effect \cite{tsui1982, laughlin1983}, it seems that a fractional Chern insulator can in principle be analogously constructed from a Chern insulator \cite{haldane1988} provided that proper electron correlation is introduced. Intriguingly, strongly correlated Chern insulators do not necessarily lead to experimentally observable fractional quantum anomalous Hall effects, because the \textcolor{black}{associated momentum-space distribution of quantum geometry}  \cite{parameswaran2012, roy2014, jackson2015} is usually not as homogeneous as that of the fractional quantum Hall effect \cite{wangjie2021}. Theoretical studies have identified the role of \textcolor{black}{the momentum-space} quantum geometry distribution in stabilizing the fractional quantum anomalous Hall effect of the strongly interacting Chern insulators \cite{parameswaran2012, roy2014, jackson2015}. This indicates that the stabilized fractional Chern insulators must be highly tunable to have appropriate topology, interaction, and quantum geometry. Fortunately, these requirements can be simultaneously satisfied in various \textcolor{black}{Moir\'e} materials \cite{ledwith2020, repellin2020, abouelkomsan2020, wilhelm2021, xieyonglong2021, spanton2018, luzhengguang2024, caijiaqi2023, zengyihang2023, park2023, xufan2023}. The first reported fractional Chern insulator is the dual-gated Bernal bilayer graphene with a rotational alignment to one of the hexagonal boron nitride-graphite gate \cite{spanton2018}. However, such a  heterostructure is not an ideal fractional \emph{Chern} insulator, because an external magnetic field as large as $30\,\text T$ must be applied to tune the topology. Magic-angle twisted bilayer graphene has the correct topology and interaction \cite{serlin2020, sharpe2019, nuckolls2020, choi2021, park2021a} but a magnetic field $\sim5\,\text T$ is required to  properly tailor the quantum geometry which stabilizes the fractional Chern insulating phase \cite{xieyonglong2021}. Remarkably, the magnitude of the required magnetic field can be greatly suppressed in multilayer graphene \cite{luzhengguang2024} and transition metal dichalcogenides \cite{caijiaqi2023, zengyihang2023, park2023, xufan2023}. Fractional quantum anomalous Hall effects have been observed in rhombohedral pentalayer graphene–hexagonal boron nitride \textcolor{black}{Moir\'e} superlattice \cite{luzhengguang2024} and twisted bilayer MoTe$_2$ \cite{caijiaqi2023, zengyihang2023, park2023, xufan2023}.

\section{QUANTUM GEOMETRY}\label{sec2}
\begin{figure*}
\centering
\includegraphics[width=140mm]{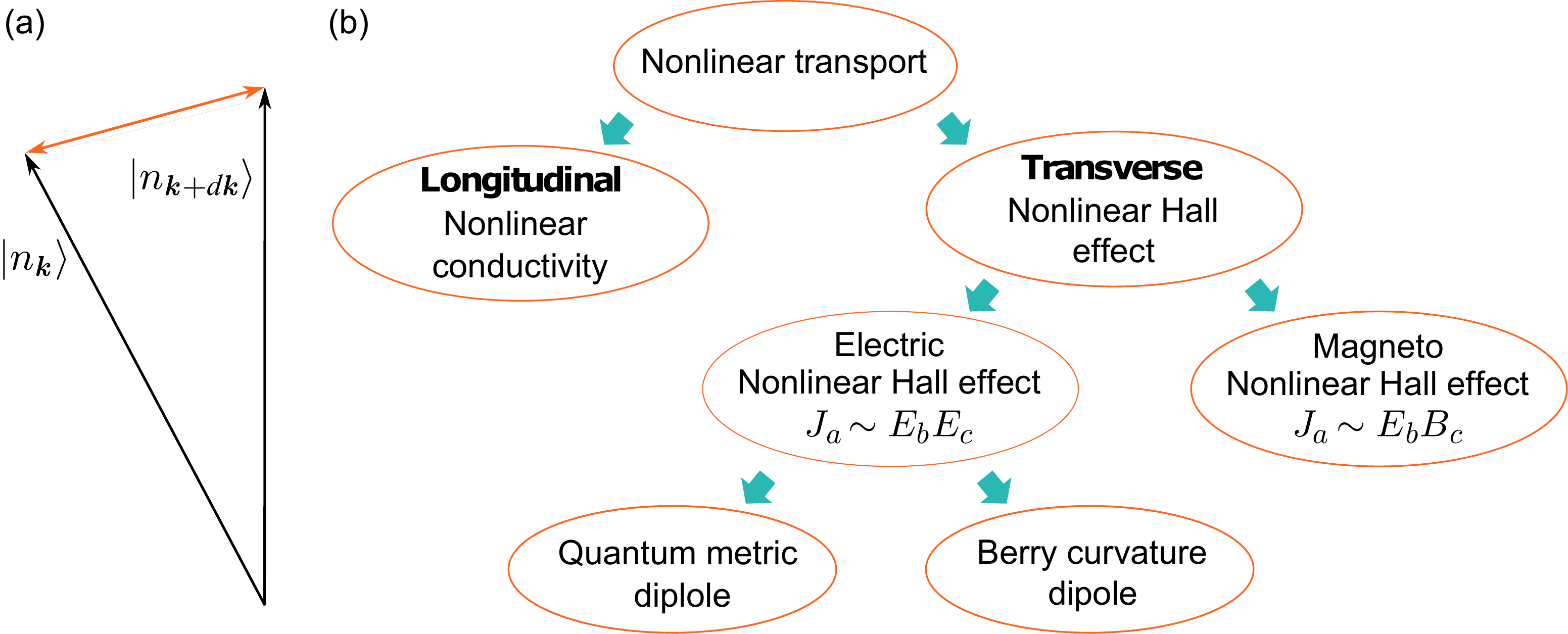}
\caption{(a) The demonstration of distance between two adjacent quantum states $|n_{\bm k+d\bm k}\rangle$ and $|n_{\bm k}\rangle$. (b) Various types of nonlinear transport associated with quantum geometry. Nonlinear longitudinal conductivity emerges from band geometry and is observable in antiferromagnetic topological insulator MnBi$_2$Te$_4$ \cite{wangnaizhou2023}. On the other hand, the nonlinear Hall effect can be categorized into two groups: (i) the electric nonlinear Hall effect; and (ii) the magneto-nonlinear Hall effect. The former may arise from the Berry curvature dipole (e.g., in WTe$_2$ \cite{maqiong2019, kangkaifei2019}, MoTe$_2$ \cite{tiwari2021, mateng2022}, WSe$_2$ \cite{huangmeizhen2023}, Cd$_3$As$_2$ \cite{shvetsov2019},  Ce$_3$Bi$_4$Pd$_3$ \cite{dzsaber2021}, and TaIrTe$_4$ \cite{kumar2021}) or the quantum metric dipole (e.g., in MnBi$_2$Te$_4$ \cite{wangnaizhou2023, gaoanyuan2023}), while the latter has been observed in kagome magnet Fe$_3$Sn$_2$ \cite{wanglujunyu2024}.    
}\label{fig1}
\end{figure*}

In the language of differential geometry, an infinitesimal distance $ds$ in a given manifold is measured by the metric $\eta^{ab}$ through
\begin{equation}\label{Eq: metric}
ds^2=\eta^{ab}dx_adx_{b},
\end{equation}
where $dx_a$ is the infinitesimal variation of the coordinate of the manifold. Formulated as a geometric theory on the Hilbert manifold, quantum mechanics allows the measurement of distance between two adjacent quantum states [respectively parameterized with momenta $\bm k+d\bm k$ and $\bm k$, see Fig.~\ref{fig1}(a)] on the $n$th band of Hamiltonian $\mathcal H$ as \cite{provost1980, torma2023}
\begin{equation}\label{Eq: dis_full}
\left||n_{\bm k +d\bm k }\rangle-|n_{\bm k }\rangle\right|^2=Q_n^{ab} dk_a dk_b,
\end{equation}
where the metric tensor is found through linearization  $|n_{\bm k +d\bm k }\rangle\simeq|n_{\bm k} \rangle+\partial_a|n_{\bm k} \rangle dk_a$ ($\partial_a\equiv\partial/\partial k_a$ is used for transparency) as 
\begin{equation} \label{Q}
Q_n^{ab}=\langle\partial_a n_{\bm k}|\partial_b n_{\bm k}\rangle.
\end{equation}
However, under a local gauge transformation $|n\rangle\rightarrow e^{i\zeta_n}|n\rangle$ ($\zeta_n$ is an arbitrary smooth function of $\bm k$ and the notation $|n\rangle\equiv|n_{\bm k}\rangle$ is adopted for transparency), the metric tensor defined in Eq.~(\ref{Q}) shows no gauge invariance
\begin{equation} 
\begin{split}
\langle\partial_a n_{\bm k}|\partial_b n_{\bm k}\rangle \rightarrow \langle\partial_a n_{\bm k}|\partial_b n_{\bm k}\rangle-\partial_a\zeta_n \mathcal A_n^b 
\\
-\partial_b\zeta_n \mathcal A_n^a+\partial_a\zeta_n \partial_b\zeta_n,
\end{split}
\end{equation}
where $\mathcal{A}_n^{a}=\langle n |i\partial_a|n \rangle$ is the intraband Berry connection \cite{berry1984, xiaodi2010}. To avoid this meaningless definition, it is worth noting that $\mathcal{A}_n^{a}$ is also gauge dependent $\mathcal{A}_n^a\rightarrow \mathcal{A}_n^a-\partial_a\zeta_n$. Consequently, a physical (i.e., gauge invariant) metric tensor, referred to as the \emph{quantum geometric tensor}, can be \textcolor{black}{naturally} defined as \cite{provost1980, torma2023, resta2011}
\begin{equation}
\mathcal{Q}_n^{ab}=\langle \partial_a n|\partial_b n\rangle-\mathcal{A}_n^a\mathcal{A}_n^b,
\end{equation}
which, by inserting the projection operator $\sum_m|m\rangle\langle m|$ to the first term, can be alternatively written as $\mathcal{Q}_n^{ab}=\sum_{m\neq n}\mathcal{A}_{nm}^a\mathcal{A}_{mn}^b$ with $\mathcal{A}_{mn}^{a}=\langle m |i\partial_a|n \rangle$ being the interband Berry connection \cite{xiaodi2010}.

The real (symmetric) part of the quantum geometric tensor reads
\begin{equation}\label{Eq: def_qm}
\begin{aligned}
\text{Re}\,\mathcal{Q}_n^{ab}=\text{Re}\langle \partial_a  n|\partial_b n\rangle-\mathcal{A}_n^{a} \mathcal{A}_n^{b} \equiv g_n^{ab},
\end{aligned}
\end{equation}
which is known as the quantum metric tensor (labelled as $g_n^{ab}$). On the other hand,  the imaginary (anti-symmetric) part of the quantum geometric tensor is
\begin{equation} \label{Eq: def_bc}
\text{Im}\,\mathcal{Q}_n^{ab}=\text{Im}\langle \partial_a n|\partial_b n\rangle=-\frac{1}{2}\Omega_n^{ab},   
\end{equation}
which is exactly the Berry curvature $\Omega_n^{ab}$ up to a constant coefficient. The quantum metric and the Berry curvature together determine the geometry of the quantum states $\mathcal{Q}_n^{ab}=g_n^{ab}-\frac{i}{2}\Omega_n^{ab}$. Furthermore, the two quantities can be related through the following inequalities \cite{roy2014, jackson2015, wangjie2021}
\begin{equation}\label{Eq: inequality}
\text{Tr}\,\bm {g}_n\ge|\boldsymbol{\Omega}_n|,\quad \text{Det}\,\bm {g}_n\ge \frac{1}{4}|\boldsymbol{\Omega}_n|^2,  
\end{equation}
\textcolor{black}{where $\bm {g}_n$ is the matrix of the quantum metric with its elements given in Eq.~\eqref{Eq: def_qm}; $\text{Tr}\,\bm {g}_n$ and $\text{Det}\,\bm {g}_n$ respectively refer to the trace and determinant of the matrix $\bm {g}_n$; $\boldsymbol{\Omega}_n$ is the Berry curvature vector with its elements determined by $\Omega_n^c=\epsilon_{abc}\Omega_n^{ab}/2$ [$\Omega_n^{ab}$ is given by Eq.~\eqref{Eq: def_bc} and $\epsilon_{abc}$ is the Levi-Civita anti-symmetric tensor].} \textcolor{black}{The saturations of the two inequalities} in Eq.~(\ref{Eq: inequality}) are usually referred to as the \emph{trace condition} and the \emph{determinant condition}, respectively.

Lastly, we remark that, by using the identity $\langle n| \partial_a \mathcal{H}|m\rangle=(\varepsilon_m-\varepsilon_n)\langle n|\partial_a m \rangle$, the quantum metric and the Berry curvature can be written in more practical forms
\begin{equation}\label{Eq: tensors}
\left\{
\begin{aligned}
g_n^{ab}&=\text{Re}\sum_{m\neq n}\frac{\langle  n|\partial_a \mathcal{H}|m\rangle\langle m| \partial_b \mathcal{H}|n \rangle}{(\varepsilon_n-\varepsilon_m)^2},\\
\Omega_n^{ab}&=-2\text{Im}\sum_{m\neq n}\frac{\langle  n|\partial_a \mathcal{H}|m\rangle\langle m| \partial_b \mathcal{H}|n \rangle}{(\varepsilon_n-\varepsilon_m)^2},
\end{aligned}
\right.
\end{equation}
which avoid the ambiguous $U(1)$ phase produced from the differentiation of the eigenstate (i.e., $|\partial_a m\rangle$).

\subsection{Two-band systems}
To better demonstrate the quantum geometric tensor, it would be instructive to provide a demo calculation of the quantum metric and the Berry curvature in some simple but non-trivial model. For this purpose, we first consider a generic two-band model
\begin{equation}
\mathcal{H}=\bm {d}\cdot\boldsymbol{\sigma},
\end{equation}
where $\boldsymbol{\sigma}=(\sigma_x,\sigma_y,\sigma_z)$ are the Pauli matrices, $\bm {d}=(d_x,d_y,d_z)$ can be arbitrary real functions of $\bm k $. The eigenenergies of the Hamiltonian $\mathcal{H}$ are $\varepsilon_\pm=\pm d$ with $d=|\bm {d}|$, corresponding to eigenstates $|\pm\rangle$. Thus, the off-diagonal matrix elements of the operator $\partial_a\mathcal{H}$ are given by $\langle \pm |\partial_a\mathcal{H} |\mp \rangle=(\partial_a d_b)\langle \pm|\sigma_b |\mp\rangle,$ where the Einstein summation convention is adopted. Making use of the identities $\langle \pm | \sigma_a | \pm \rangle=d_a/d$ and $\sigma_a\sigma_b = i\epsilon_{abc}\sigma_c +\delta_{ab}$, where $\delta_{ab}$ is the Kronecker symbol, it is straightforward to find $\langle \pm |\sigma_a |\mp\rangle \langle \mp|\sigma_b |\pm \rangle=\delta_{ab}-{d_ad_b}/{d^2}\pm i\epsilon_{abc}d_c/d$. Consequently, Eq. (\ref{Eq: tensors}) is simplified to
\begin{equation}\label{Eq: 2-band}
\left\{
\begin{aligned}
g_\pm^{ab}&=\frac{1}{4d^2}\Big[\partial_a \bm {d}\cdot\partial_b \bm {d}-\frac{1}{d^2}(\partial_a\bm {d}\cdot\bm {d})(\partial_b\bm {d}\cdot\bm {d})\Big],\\
\Omega_\pm^{ab}&=\mp\frac{(\partial_a\bm {d} \times\partial_b\bm {d})\cdot \bm {d}}{2d^3},\\
\end{aligned}
\right.        
\end{equation}
which are applicable for any two-band systems. The treatment for a generic $n$-band system can be found in  Ref. \cite{graf2021}. 

As a concrete example, we consider a two-dimensional massive Dirac model 
\begin{equation}
\mathcal{H}=vk_x\sigma_x+vk_x\sigma_y+m\sigma_z,
\end{equation}
where velocity $v$ and mass $m$ are model parameters. The vector $\bm {d}$ is given by $\bm {d}=(vk_x,vk_x,m)$ with $d=\sqrt{v^2(k_x^2+k_y^2)+m^2}$. According to Eq. (\ref{Eq: 2-band}), the quantum metric tensor and the Berry curvature tensor respectively read
\begin{equation}
\begin{split}
\bm {g}_{\pm} &=\frac{v^2}{4d^4}
\begin{bmatrix}
v^2k_y^2+m^2& -v^2k_xk_y\\
-v^2k_xk_y& v^2k_x^2+m^2
\end{bmatrix},
\\
\bm\Omega_\pm &=\frac{v^2}{4d^4}\begin{bmatrix}
0 & \mp 2dm
\\
\pm 2dm & 0
\end{bmatrix},
\end{split}
\end{equation}
\textcolor{black}{which lead to $\text{Tr}\,\bm g_\pm=g_\pm^{xx}+g_\pm^{yy}=\tfrac{v^2(d^2+m^2)}{4d^2}$, $\text{Det}\,\bm g_\pm=g_\pm^{xx}g_\pm^{yy}-g_\pm^{xy}g_\pm^{yx}=\tfrac{m^2v^4}{16d^6}$, and $|\bm \Omega_\pm|=|\tfrac 1 2\Omega_\pm^{xy}-\tfrac 1 2 \Omega_\pm^{yx}|=\tfrac{|m|v^2}{2d^3}$. It is straightforward to check that both inequalities in Eq.~\eqref{Eq: inequality} stand.}

\section{NONLINEAR TRANSPORT}\label{sec3}
\textcolor{black}{The determinants of quantum geometry (i.e., band dispersions and wave functions) coincide with the inputs of the Boltzmann transport equation. This fact implies profound impacts on transport exerted by quantum geometry. In particular, quantum geometry has been known to produce} various types of nonlinear transport [Fig.~\ref{fig1}(b)] in both transverse \cite{sodemann2015, duzz2018, duzz2019, maqiong2019, kangkaifei2019, huangmeizhen2023, tiwari2021, mateng2022, shvetsov2019, kumar2021, dzsaber2021, gaoyang2014, kaplan2024, wangchong2021, liuhuiying2021, gaoanyuan2023, wangnaizhou2023, wanglujunyu2024} and longitudinal directions \cite{kaplan2024, wangnaizhou2023}. The former can be further classified into two categories: (i) the electric nonlinear Hall effect arising from either the Berry curvature dipole \cite{sodemann2015, duzz2018, duzz2019, maqiong2019, kangkaifei2019, huangmeizhen2023, tiwari2021, mateng2022, shvetsov2019, kumar2021, dzsaber2021} or the quantum metric dipole \cite{gaoyang2014, kaplan2024, wangchong2021, liuhuiying2021, gaoanyuan2023, wangnaizhou2023}; and (ii) the magneto-nonlinear Hall effect \cite{gaoyang2014, wanglujunyu2024}. In this section, we will review these different types of nonlinear transport resulting from quantum geometry.

\subsection{Electric nonlinear Hall effect}
Formulating the Hall voltage as a transverse current $\bm J$ and the driving current as the applied electric field $\bm E$, the transport can be formally expressed as a power series
\begin{equation} \label{Je}
J_a=\sigma_{ab}E_b+\sigma_{abc}E_bE_c,
\end{equation}
where the tensor $\sigma_{ab}$ is the linear Hall conductivity (for $a\neq b$) and $\sigma_{abc}$ is the (second-order) nonlinear Hall conductivity (for nonidentical indices $a$, $b$, and $c$). For simplicity, we here do not go to the cubic regime and beyond, which should be typically less dominating. The appearance of linear/nonlinear Hall response is subject to the symmetry and quantum geometry \cite{sodemann2015, duzz2018, duzz2019, maqiong2019, kangkaifei2019, huangmeizhen2023, tiwari2021, mateng2022, shvetsov2019, kumar2021, dzsaber2021, gaoyang2014, kaplan2024, wangchong2021, liuhuiying2021, gaoanyuan2023, wangnaizhou2023}.

The electric current can be explicitly expressed as
\begin{equation}\label{Eq: e-Current}
\begin{aligned}
J_a=-e\sum_n\int [d\bm k ]v_n^a f,\\
\end{aligned} 
\end{equation}
where $[d\bm k ]$ represents $d^d\bm k /(2\pi)^d$ with dimension $d$, $v^a_n$ is the electron velocity for the $n$th band, $f$ is the non-equilibrium distribution function. In the absence of magnetic field, $f$ can be determined by the Boltzmann equation
\begin{equation}
\frac{-\tau e}{\hbar}\bm {E}\cdot\partial_{\bm k}  f=f_0-f, 
\end{equation}
where $\tau$ is the relaxation time, and $f_0$ is the Fermi-Dirac distribution. The solution of Boltzmann equation can be expressed as a power series of $\bm {E}$ as
\begin{equation}
\begin{aligned}
f=\sum_n \left( \frac{\tau e}{\hbar}\bm {E}\cdot \partial_{\bm k} \right)^n f_0.
\end{aligned}     
\end{equation}
According to the wave packet dynamics \cite{xiaodi2010,gaoyang2014,GaoY15prb,liuhuiying2021,wangchong2021,liuhuiying2022,WangJ23prb} and Luttinger-Kohn method \cite{kaplan2024}, up to the second-order of electric field, the velocity can be written as
\begin{equation}
\begin{aligned}
v_n^a=&\frac{1}{\hbar}\partial_{a}\varepsilon_n+\frac{e}{2\hbar}(\Omega_n^{ab}E_b+\Omega_n^{ac}E_c)\\
&+\frac{e}{2\hbar}\big[3\partial_{a}G_n^{bc}-(\partial_cG_n^{ab}+\partial_bG_n^{ac})\big]E_bE_c, 
\end{aligned}    
\end{equation}
where $G_n^{ab}=2e\text{Re}\sum_{m\neq n}\mathcal{A}_{nm}^a\mathcal{A}_{mn}^{b}/(\varepsilon_n-\varepsilon_m)$ is the Berry connection polarizability related to the quantum metric through $G_n^{ab}=-e\partial g_n^{ab}/\partial\varepsilon_n$ \cite{gaoyang2014,GaoY15prb}. Upon plugging $v_n^a$ into Eq.~(\ref{Eq: e-Current}), the nonlinear conductivity reads
\begin{equation} \label{NL_cond}
\begin{aligned}
\sigma_{abc}=&-\frac{e^3\tau^2}{\hbar^3}\sum_n\int [d\bm k ] \big(\partial_a\partial_b  \partial_c\varepsilon_n\big)f_0\\
&+\frac{e^3\tau}{2\hbar^2}\sum_n\int [d\bm k ]\big(\partial_c\Omega_n^{ab}+\partial_b\Omega_n^{ac}\big)f_0\\
&-\frac{e^2}{2\hbar}\sum_n\int [d\bm k ]\big[2\partial_aG_n^{bc}-\\
&\big(\partial_cG_n^{ab}+\partial_bG_n^{ac}\big)\big]f_0,
\end{aligned} 
\end{equation}
where the first term is referred to as the nonlinear Drude weight \cite{kaplan2024} and the second (third) term is associated with the Berry curvature dipole \cite{sodemann2015,duzz2018} (Berry connection polarizability dipole, also known as the band-normalized quantum metric dipole \cite{gaoyang2014,liuhuiying2021,wangchong2021}).

\subsubsection{Berry curvature dipole induced nonlinear Hall effect}
In the presence of time-reversal symmetry $\mathcal T$, the linear Hall response is prohibited. With inversion symmetry $\mathcal P$ broken, the Berry curvature dipole [second term, Eq.~(\ref{NL_cond})] becomes the major source of transport and results in an electric nonlinear Hall effect \cite{sodemann2015, duzz2018, duzz2019}.

\begin{figure}[t]
\centering
\includegraphics[width=65mm]{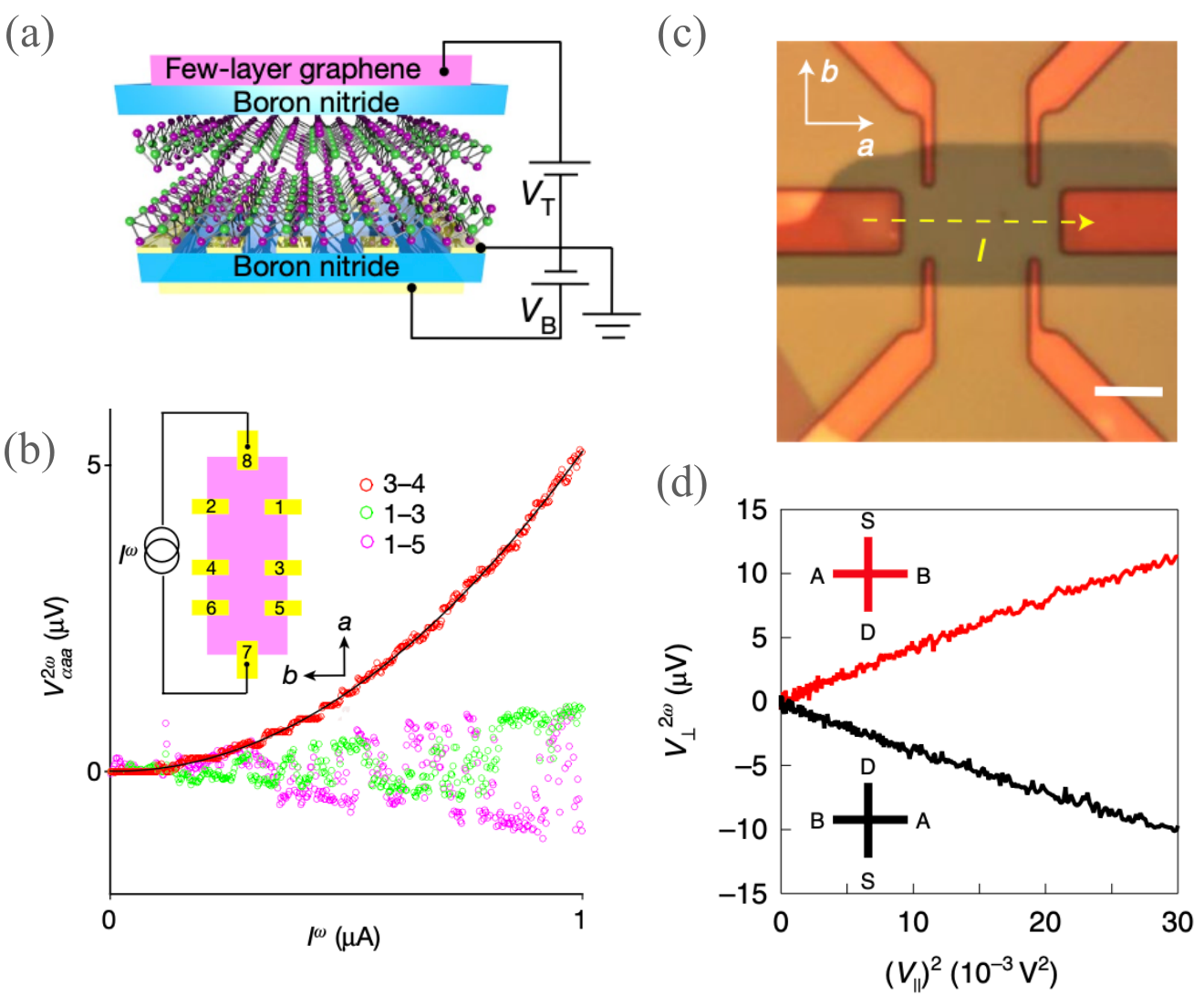}
\caption{Electric nonlinear Hall effect in WTe$_2$. (a) Schematic plot of a dual-gated bilayer WTe$_2$ device encapsulated in hexagonal boron nitride electrodes. (b) Nonlinear Hall (red) and longitudinal (green, purple) voltages $V^{2\omega}$ versus the driving current $I^{\omega}$. Panels (a, b) are adapted from Ref.~\cite{maqiong2019}. (c) Optical image of a Hall bar device of few-layer WTe$_2$. (d) Nonlinear Hall voltages. The driving current flows from source (S) to drain (D) and voltages are measured between electrodes A and B. Panels (c, d) are adapted from Ref.~\cite{kangkaifei2019}.
}\label{fig2}
\end{figure}

This Berry curvature dipole induced nonlinear Hall effect has been proposed in various transition metal dichalcogenides and semimetals. The first experimental realizations of the nonlinear Hall effect adopt dual-gated few layer WTe$_2$ Hall bar devices \cite{maqiong2019, kangkaifei2019}, as illustrated in Figs.~\ref{fig2}(a)-(d). The effect has also been observed in MoTe$_2$ \cite{tiwari2021, mateng2022}, strained and twisted WSe$_2$ \cite{huangmeizhen2023}, Dirac semimetal Cd$_3$As$_2$ \cite{shvetsov2019}, Weyl-Kondo semiemtals Ce$_3$Bi$_4$Pd$_3$ \cite{dzsaber2021}, and Weyl semimetal TaIrTe$_4$ \cite{kumar2021}.

\subsubsection{Quantum metric dipole induced nonlinear Hall effect}
We now consider a system with both time-reversal and inversion symmetries broken but $\mathcal{PT}$ symmetry preserved. The linear Hall effect and electric nonlinear Hall effect arising from the Berry curvature dipole [second term, Eq.~(\ref{NL_cond})] are both prohibited. However, the electric nonlinear Hall effect itself can possibly survive \cite{gaoanyuan2023, wangnaizhou2023, gaoyang2014, kaplan2024, wangchong2021, liuhuiying2021}, contributed by the nonlinear Drude weight [first term, Eq.~(\ref{NL_cond})] and quantum metric dipole [third term, Eq.~(\ref{NL_cond})]. The different $\tau$ dependence allows us to distinguish the two contributions with a simple scaling law \cite{kaplan2024, gaoanyuan2023, wangnaizhou2023}
\begin{equation} \label{scaling}
\chi=\eta_2\sigma_\parallel^2+\eta_0,
\end{equation}
where the first and second terms respectively correspond to the nonlinear Drude weight and the quantum metric dipole; and $\sigma_\parallel\sim\tau$ is the linear longitudinal conductivity.

\begin{figure*}[!t]
\centering
\includegraphics[width=140mm]{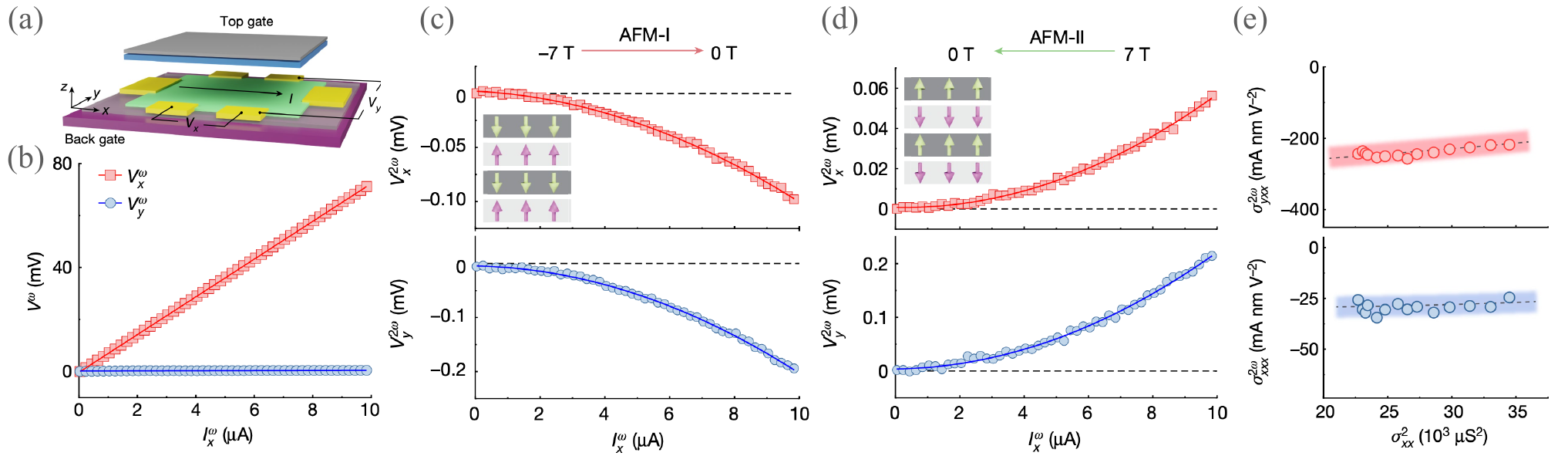}
\caption{Nonlinear Hall effect and longitudinal conductivity. (a) Schematic plot of a 6-end Hall bar device of 4 \textcolor{black}{septuple layer} MnBi$_2$Te$_4$ (green) with a top gate (grey) and a back gate (purple). An AC driving current $I^\omega$ is applied along the $x$ direction, while linear/nonlinear voltages are measured along $x$ and $y$ directions using the standard lock-in technique. (b) Linear longitudinal (red) and Hall (blue) voltages. (c, d) Nonlinear longitudinal (red) and Hall (blue) voltages for the AFM-I and AFM-II phases of MnBi$_2$Te$_4$. Insets illustrate the magnetization of each septuple layer in AFM-I and AFM-II phases. (e) Scaling of the nonlinear Hall (red) and longitudinal (blue) conductivity with respect to the square of the linear longitudinal conductivity. The dashed lines are the linear fits according to Eq.~(\ref{scaling}). All panels are adapted from Ref.~\cite{wangnaizhou2023}.
}
\label{fig3}
\end{figure*}

To experimentally observe the electric nonlinear Hall effect, we first require a platform breaking both $\mathcal P$ and $\mathcal T$ but preserving $\mathcal{PT}$. Such symmetry requirements could be naturally satisfied in antiferromagnets \cite{kaplan2024, wangchong2021, liuhuiying2021} such as MnBi$_2$Te$_4$. MnBi$_2$Te$_4$ is a magnetic topological insulator with adjacent ferromagnetic layers exhibiting opposite magnetization, and thus odd- (even-) layered MnBi$_2$Te$_4$ realizes ferromagnetic (antiferromagnetic) topological insulators. For a dual-gated 4 \textcolor{black}{septuple layer} MnBi$_2$Te$_4$ Hall bar device [Fig.~\ref{fig3}(a)], two types of N\'eel order, referred to as AFM-I and AFM-II [Insets, Fig.~\ref{fig3}(c, d)], can be realized by respectively sweeping the magnetic field from $-7\,\text T$ to $0\,\text T$ and $7\,\text T$ to $0\,\text T$ \cite{wangnaizhou2023}. Applying an ac driving current $I^\omega$ in the $x$ direction [Fig.~\ref{fig3}(a)], both linear and nonlinear Hall responses can be measured simultaneously through the standard lock-in technique. In the linear regime, both AFM-I and AFM-II exhibits longitudinal response $V_y^\omega$ but vanishing Hall voltage $V_x^\omega$, indicating no net magnetization and thus confirming the N\'eel order [Fig.~\ref{fig3}(b)]. On the other hand, pronounced nonlinear Hall voltages $V_y^{2\omega}$ are observed for both AFM-I and AFM-II except for a sign difference [Figs.~\ref{fig3}(c, d)], which suggests no contribution from the Berry curvature dipole \cite{gaoanyuan2023}. Such electric nonlinear Hall effect persists up to the N\'eel temperature $T_{\text{N\'eel}}\sim20\,\text{K}$. To distinguish the contribution from the nonlinear Drude weight and the quantum metric dipole, a scaling of the nonlinear Hall conductivity [Eq.~(\ref{scaling})] is conducted in the regime $T=2\sim10\,\text{K}$, in which the carrier density is nearly a constant and thus $\sigma_{xx}^\omega$ is approximately proportional to $\tau$. The $\sigma_{yxx}^{2\omega}$ versus $(\sigma_{xx}^{\omega})^2$ curve exhibits linear behavior with the intercept (slope) associated with the quantum metric dipole (Drude weight). The quantum metric dipole is found to dominate the nonlinear transport, which is justified by the flatness of the curve [Fig.~\ref{fig3}(e)].

\subsection{Magneto-nonlinear Hall effect}
In the presence of purely electric fields, the nonlinear Hall effect is also purely electric and arises from either the Berry curvature dipole \cite{sodemann2015, duzz2019, duzz2018, maqiong2019, kangkaifei2019, huangmeizhen2023, tiwari2021, mateng2022, shvetsov2019, kumar2021, dzsaber2021} or the quantum metric dipole \cite{gaoanyuan2023, wangnaizhou2023, gaoyang2014, kaplan2024, wangchong2021, liuhuiying2021}. With the participation of magnetic fields, the nonlinear Hall family can be further expanded to the magneto-paradigm. The response reads \cite{gaoyang2014}
\begin{equation} \label{Jm}
J_a=\sigma_{ab}E_b+\chi_{abc}E_bB_c,
\end{equation}
where the first term is the ordinary Hall response (for $a\neq b$) and the second bilinear term represents the magnetoelectric response with $\chi_{abc}$ being the magneto-nonlinear Hall conductivity (for nonidentical indices $a$, $b$, and $c$). For simplicity, the purely electric nonlinear response (i.e., $\sigma_{abc}E_bE_c$) is neglected in Eq.~(\ref{Jm}). The terms associated with cubic or higher combinations of electric and magnetic fields are also neglected. 

The bilinear term in Eq.~(\ref{Jm}) is originated from quantum geometry \cite{wanglujunyu2024}. Such a term can be intuitively understood as the additional Berry connection induced by a magnetic field \cite{gaoyang2014} in contrast to the extra Berry connection resulting from the electric field in the electric nonlinear Hall effect. Explicitly, this additional Berry connection reads \cite{gaoyang2014, wanghui2024, wanglujunyu2024}
\begin{equation} \label{Am}
\textcolor{black}{
\mathcal A_n^{(1),b}=B_a[F_n^{\text{S},ab}+F_n^{\text{O},ab}],
}
\end{equation}
where $F_n^{\text{S}(\text{O}),ab}$ is the anomalous spin (orbital) polarizability. Explicitly, they are expressed as
\begin{align}
F_n^{\text{S},ab} =&-2\text{Re}\sum_{m\neq n}\frac{\mathcal M_{nm}^{\text S, a} \mathcal A_{mn}^b}{\varepsilon_n-\varepsilon_m},
\\
\begin{split}
F_n^{\text{O},ab} =&-2\text{Re}\sum_{m\neq n}\frac{\mathcal M_{nm}^{\text O, a} \mathcal A_{mn}^b}{\varepsilon_n-\varepsilon_m} 
\\
&- \frac{e}{2\hbar}\epsilon_{acd}\partial_cg_n^{db},
\end{split}
\end{align}
where $\mathcal A _{mn}^a$ is the interband Berry connection in the absence of applied electromagnetic fields; $\varepsilon_n$ is the dispersion of the $n$th band; $\bm {\mathcal M}_{mn}^{\text{S}}=-g\mu_B\bm s_{mn}$ is the interband spin magnetic moment with $\bm s_{mn}$ being the matrix element of the spin operator, $\mu_B$ being the Bohr magneton, and $g$ being the g-factor \cite{wanghui2024}; $\bm {\mathcal M}_{mn}^{\text{O}}=e/2\sum_{l\neq n}(\bm v_{ml}+\delta_{lm}\bm v_n)\times \bm {\mathcal A}_{ln}/2$ is the interband orbital magnetic moment with $\bm v_{mn}$ being the matrix element of the velocity operator \cite{wanghui2024}; and $g_n^{ab}$ is the quantum metric tensor. In the presence of a magnetic field, the dispersion of the $n$th band is renormalized to
\begin{equation}
\tilde{\varepsilon}_n=\varepsilon_n-\bm B \cdot (\bm{\mathcal M}_n^\text{S}+\bm{\mathcal M}_n^\text{O}),
\end{equation}
where $\bm{\mathcal M}_n^\text{S(O)}$  \cite{xiaodi2010} is the intraband spin and orbital moment. Consequently, the electron velocity up to the quadratic order of electromagnetic field reads
\begin{equation} \label{v_op}
\tilde{v}_n^{a}=\textcolor{black}{\frac 1 \hbar} \partial_a\tilde{\varepsilon}_n-\epsilon_{abc}\frac{e}{\hbar}E_b\left[\Omega_n^{c}+\Omega_n^{(1),c}\right],
\end{equation}
where $\bm\Omega_n$ is the unperturbed Berry curvature and $\bm\Omega_n^{(1)}=\nabla_{\bm k} \times\mathcal A_n^{(1)}$ is the field induced Berry curvature [see Eq.~(\ref{Am})]. Plugging Eq.~(\ref{v_op}) and the distribution function $f_0(\tilde{\varepsilon}_n)$ into Eq.~(\ref{Eq: e-Current}), the magneto-nonlinear Hall conducticity can be solved as
\begin{equation} \label{chi_m}
\chi_{abc}=\frac{e^2}{\hbar}\sum_n\int[d\bm k] \left[ \Theta_n^{\text{S},abc} + \Theta_n^{\text{O},abc} \right]f_0'
\end{equation}
where \textcolor{black}{$\Theta_n^{\text{S}(\text{O}),abc}=\hbar v_n^bF_n^{\text{S}(\text{O}),ca}-\hbar v_n^aF_n^{\text{S}(\text{O}),cb}-\varepsilon_{abd}\Omega_n^d\mathcal M_n^{\text{S}(\text{O}),c}$.} Analogous to the Berry curvature dipole and quantum metric dipole, the first two terms in $\Theta_{abc}^{\text{S}(\text{O})}$ may be referred to as the anomalous spin (orbital) polarizability dipole, reflecting the band geometric nature of the magneto-nonlinear Hall effect. The appearance of $f_0'$ in Eq.~(\ref{chi_m}) indicates that the magneto-nonlinear Hall conductivity is a Fermi surface property.

Unlike the detection of the electric nonlinear Hall effect which requires platforms of appropriate symmetries ($\mathcal P$-breaking but $\mathcal T$-preserving for the Berry curvature dipole \cite{maqiong2019, kangkaifei2019, huangmeizhen2023, tiwari2021, mateng2022, shvetsov2019, kumar2021, dzsaber2021}, while $\mathcal P$ and $\mathcal T$ respectively broken but $\mathcal {PT}$-preserving for the quantum metric dipole \cite{gaoanyuan2023, wangnaizhou2023}), there are in general no symmetry requirements for the observation of the magneto-nonlinear Hall effect \cite{wanglujunyu2024}. However, the scaling of the magneto-nonlinear Hall effect is $EB$ [Eq.~(\ref{Jm})], identical to that of the ordinary Hall effect resulting from the Lorentz force \cite{gaoyang2014}.  Therefore, one would prefer an in-plane magnetic field, with which the ordinary Hall effect is suppressed.

\begin{figure}[t!]
\centering
\includegraphics[width=65mm]{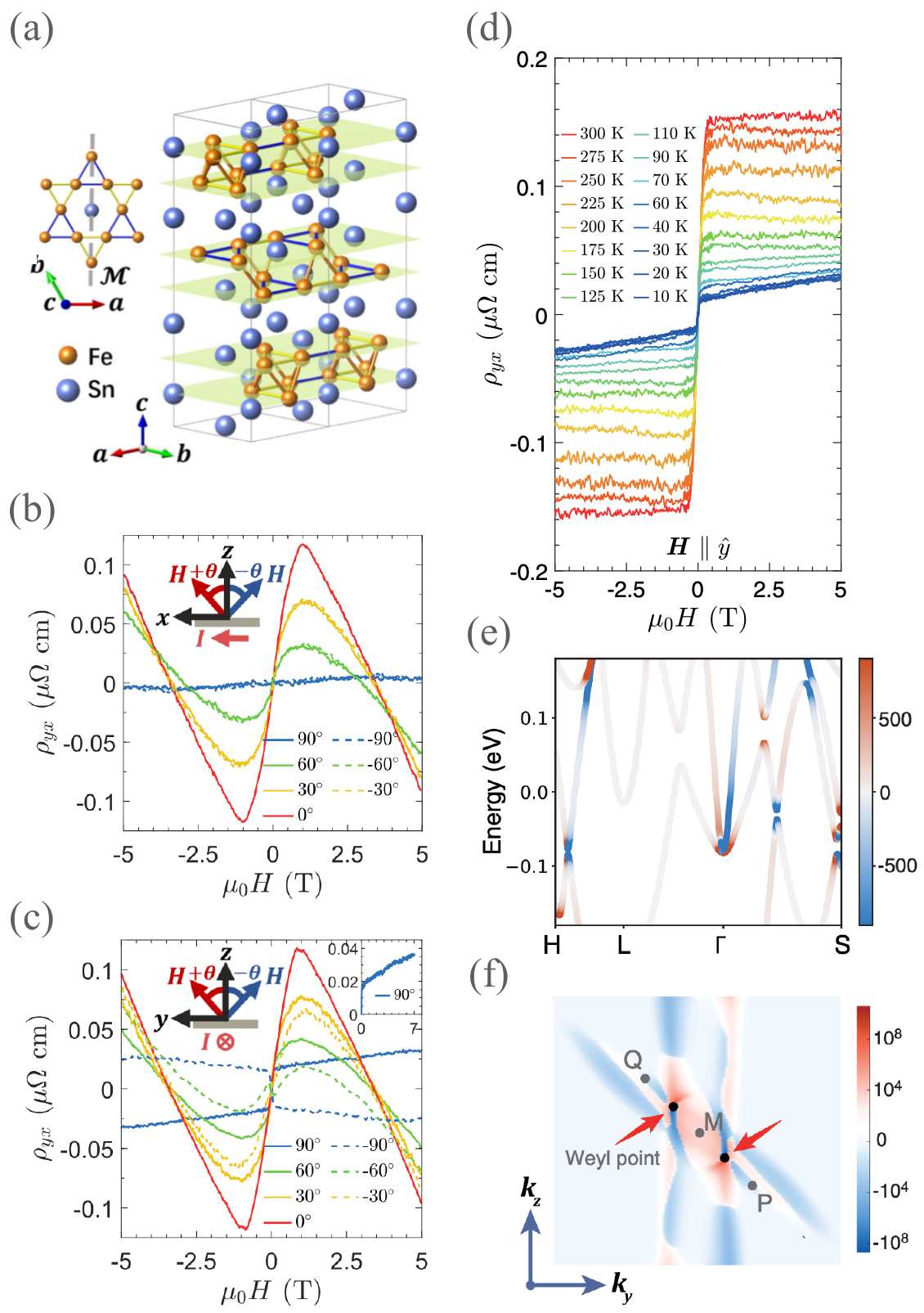}
\caption{(a) Schematic atomic structure of Fe$_3$Sn$_2$ with gold (blue) balls representing iron (stannum) atoms. (b) Angle-resolved Hall resistivity for a magnetic field in the $z$-$x$ plane. (c) Angle-resolved Hall resistivity for a magnetic field in the $y$-$z$ plane. Inset: In-plane Hall resistivity occurs with a $y$-direction magnetic field. (d) Magnetic field dependence of Hall resistivity at different temperatures. (e) The $yy$ component of the anomalous orbital polarizability projected on the first-principles band structure. (f) Distribution of the anomalous orbital polarizability dipole on the Fermi surface, where a pair of Weyl points serve as the hot spots. All panels are adapted from Ref.~\cite{wanglujunyu2024}.
}
\label{fig4}
\end{figure}

One ideal platform for the observation of the magneto-nonlinear Hall effect is the ferromagnetic semimetal Fe$_3$Sn$_2$, because the ferromagnetic order makes the anomalous orbital polarizability the dominant contribution to the conductivity [Eq.~(\ref{chi_m})] near the band degeneracy \cite{wanglujunyu2024}. Fe$_3$Sn$_2$ comprises of kagome Fe$_3$Sn layers and honeycomb Sn layers stacked along the $c$ axis [Fig.~\ref{fig4}(a)] and exhibits a mirror plane, labelled as $M_x$, perpendicular to the $a$ axis. For a tilted magnetic field with an angle $\theta$ with respect to the $z$ in the $xz$ plane, one finds that the Hall resistivity $\rho_{yx}$ remains intact under angle inversion $\theta\rightarrow-\theta$ and completely vanishes along the $x$ direction (i.e., $\theta=90^\circ$) [Fig.~\ref{fig4}(b)]. Such behaviors imply that only the out-of-plane component of the magnetic field plays an role, consistent with the ordinary Hall effect. By contrast, when the magnetic field is tilted in the $yz$ plane, $\rho_{yx}$ is not invariant under $\theta\rightarrow-\theta$, ruling out the possibility that the ordinary Hall effect acts as the sole resource [Fig.~\ref{fig4}(c)]. Remarkably, $\rho_{yx}$ remains finite for $\theta=90^\circ$ and thus implies an in-plane Hall effect, which is prohibited for an $x$-direction magnetic field because of the mirror symmetry $M_x$. The Hall resistivity of such an in-plane Hall effect exhibits a sudden jump at zero field and linear-in-$H$ behavior for finite magnetic fields, i.e., $\rho_{yx}=\rho_{\text{IAHE}}^0+\rho_{\text{IAHE}}^H$ [Fig.~\ref{fig4}(d)], where the former arises from the magnetization and the latter is related to the spin/orbital coupling of carriers to the magnetic field \cite{wanglujunyu2024}. The conductivity associated the second term is $\sigma_{\text{IAHE}}^H\simeq -\rho_{\text{IAHE}}^H/\rho_{xx}^2$], which is also linear in $H$ and thus implies a magneto-nonlinear Hall effect.

First-principles calculations have confirmed that the observed in-plane magneto-nonlinear Hall effect mainly arises from the dipole of anomalous orbital polarizability [see Eq.~(\ref{chi_m})], which dominates other contributions by at least one order in magnitude \cite{wanglujunyu2024}. The anomalous orbital polarizability is found to be most pronounced at  band degeneracies or around small band gaps [Fig.~\ref{fig4}(e)], among which those close to the Fermi surface will significantly contribute to the magneto-nonlinear Hall conductivity [Eq.~(\ref{chi_m})]. The relevant band structures comprise four pairs of Weyl points, one of which is located on the Fermi surface and connected by the $M_x\mathcal T$ symmetry. The two Weyl points serve as hot spots of the anomalous orbital polarizability [Fig.~\ref{fig4}(f)].

\subsection{Nonlinear longitudinal conductivity}
\textcolor{black}{Besides the electric and magneto-nonlinear Hall effects, quantum geometry is also predicted to be responsible for the nonlinear longitudinal transport through a perturbative approach \cite{kaplan2024}. The existence of quantum geometry induced nonlinear longitudinal transport is also justified with density matrix calculations \cite{das2023}, though the obtained nonlinear conductivity is quantitatively different from the perturbative predication. Controversially, nonlinear longitudinal transport seems prohibited in the framework of wave packet dynamics, if the electric field correction to the Fermi-Dirac distribution is also enclosed \cite{gaoyang2014}.
}

The nonlinear transport in the longitudinal direction has been experimentally observed in MnBi$_2$Te$_4$ with the same dual-gated Hall bar device \cite{wangnaizhou2023} [Fig.~\ref{fig3}(a)]. For both AFM-I and AFM-II phases, pronounced longitudinal voltages of opposite signs are observed [Figs.~\ref{fig3}(c, d)]. Same as the Hall voltages, the observed longitudinal responses persist up to the N\'eel temperature $T_{\text{N\'eel}}\sim20\,\text K$. The scaling of the longitudinal conductivity reveals a linear dependence of $\sigma_{xxx}$ on $(\sigma_{xx}^\omega)^2$ [Fig.~\ref{fig3}(e)], \textcolor{black}{implying the presence of quantum metric dipole contribution.}

\section{\textcolor{black}{FLAT-BAND SUPERCONDUCTORS}}\label{sec4}
Understanding the properties of unconventional superconductors is beyond the celebrated Bardeen-Cooper-Schrieffer formalism and promotes a great variety of theories \cite{stewart2017}. Amongst them, theories based on quantum geometry have recently attracted great attentions for the explanation of the enhanced transition temperature in flat-band superconductors \cite{peotta2015, julku2016, lianglong2017, torma2018, huhtinen2022}. Because of the band flatness, one may naively expect the associated inert electrons to only support a vanishingly small supercurrent. However, a sizable supercurrent has been recently observed in twisted bilayer graphene \cite{tianhaidong2023}, a well recognized two-dimensional flat-band superconductor \cite{caoyuan2018a}. It is thus critically important to understand the origination of this supercurrent.

\begin{figure*}[!t]
\centering
\includegraphics[width=140mm]{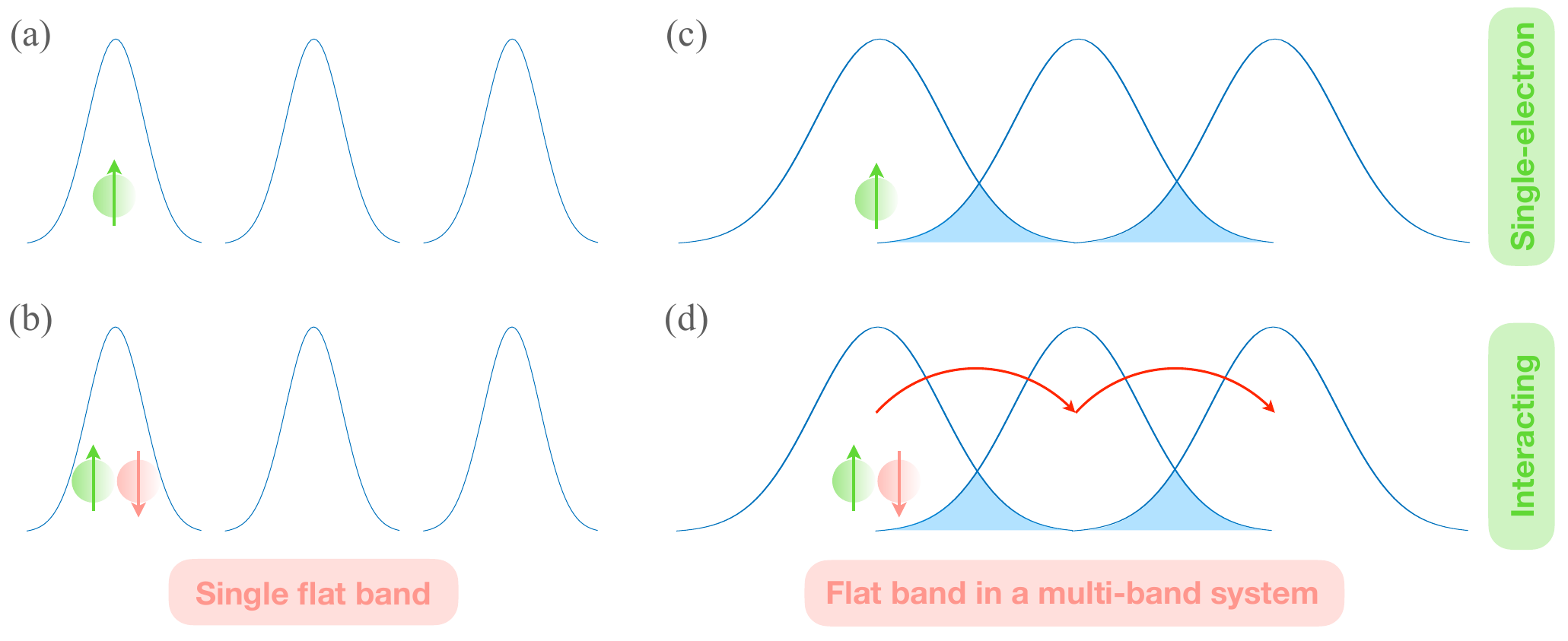}
\caption{Schematic plot of Wannier function overlap. (a) A single flat band formed by well separated Wannier functions in the absence of interactions. (b) A single flat band robust against interaction which does not create sizable Wannier function overlap. (c) A flat band in a multi-band system formed by destructive interference of Wannier functions. (d) Interaction interrupts the destructive interference and causes a supercurrent. Reproduced with changes from Ref.~\cite{peotta2023}.
}
\label{fig5}
\end{figure*}

{\color{black}
\subsection{Quantum geometric theory}
The emergence of flat bands in superconductors can be understood in the view of Wannier function overlap, which is governed by quantum geometry. A single flat band can possibly arise from well isolated Wannier functions [Fig.~\ref{fig5}(a)] and can stay robust against interaction, provided that the interaction does not create sizable overlap between adjacent Wannier functions [Fig.~\ref{fig5}(b)]. Consequently, there is no transport on the single flat band. Alternatively, a flat band can emerge in a multi-band system, where Wannier functions interfere destructively [Fig.~\ref{fig5}(c)]. Interaction can interrupt the interference and causes a supercurrent [Fig.~\ref{fig5}(d)]. A more theoretical justification on the absence (presence) of supercurrents in single- (multi-) band systems is presented in the following.
}

\subsubsection{Single-band model}
We first consider a toy model with a single isotropic parabolic band $\varepsilon =\hbar^2k^2/2m_{\text{eff}}$. In the large effective mass limit (i.e., $m_\text{eff}\rightarrow \infty$), the band become flattened.

The supercurrent of the model in the London gauge reads \cite{tinkham2004}
\begin{equation} \label{london}
J_a=-D^{ab} A_b,
\end{equation}
where $A_b$ is the applied magnetic field and $D^{ab}$ is the superfluid weight tensor defined as \cite{peotta2015}
\begin{equation} \label{D_sing}
D^{ab}=\frac{e^2}{\hbar^2}\int[d\bm k] (\partial_a\partial_b\varepsilon)f_0.
\end{equation}
It is straightforward to check that the supercurrent $\bm J_s=({ne^2}/{m_\text{eff}}) \bm A$ indeed vanishes when the single band is flattened in the $m_\text{eff}\rightarrow \infty$ limit.

\subsubsection{Multi-band model}
The fact that a single flat band is indeed insufficient to carry a supercurrent does not contradict with the observed flat-band superconductors \cite{tianhaidong2023}, whose characterization requires a more complicated multi-band model, generically written as 
\begin{align} \label{hubbard}
\mathcal{H}&=\sum_{i, j, \sigma}t_{i,j}c_{i,\sigma}^\dagger c_{j,\sigma}-U\sum_{i} n_{i,\uparrow} n_{i,\downarrow}
\\
&\simeq\sum_{i, j, \sigma}t_{i,j}c_{i,\sigma}^\dagger c_{j,\sigma} + \sum_{i} \left( \Delta_i c_{i,\uparrow}^\dagger c_{i,\downarrow}^\dagger + \text{H.c.}  \right) \nonumber,
\end{align}
where $c_{i, \sigma}$ annihilates a spin-$\sigma$ electron on the $i$th site; $t_{i,j}$ is the spin-independent hopping; $U>0$ is the onsite Hubbard interaction; and $\Delta_i=-U\langle c_{i,\downarrow} c_{i,\uparrow} \rangle$ is the superconductor order parameter at the mean-field level.

In the presence of time-reversal symmetry, the solution of the two-body problem associated with the Hubbard model [Eq.~(\ref{hubbard})] estimate the effective mass of the bound states as \cite{peotta2015, julku2016, lianglong2017, torma2018, huhtinen2022}
\begin{equation} \label{m_eff}
\left[\frac{1}{m_\text{eff}}\right]_{ab}=\frac{U \mathcal{V}}{N\hbar^2} \int[d\bm k] g_n^{ab} 
\end{equation}
where $ \mathcal{V}$ is the volume of the unit cell; $N$ is the number of the sites in a unit cell; and $g_n^{ab} $ is the quantum metric tensor. Equation~(\ref{m_eff}) suggests a finite mass in the presence of a nonzero quantum metric tensor, thus justifying the existence of the supercurrent. To explicitly quantify the supecurrent [cf., Eq.~(\ref{london})], the superfluid weight has to be evaluated. Assuming a real and spatially uniform order parameter $\Delta_i\equiv\Delta$, the superfluid weight 
\begin{equation} \label{sc_weight}
D^{ab}=D_{\text{conv}}^{ab}+D_{\text{geom}}^{ab},
\end{equation}
is found to have a conventional part $D_{\text{conv}}^{ab}$ and a geometric part $D_{\text{geom}}^{ab}$ \cite{peotta2015, lianglong2017}, written as
\begin{align} 
D_{\text{conv}}^{ab}&=\frac{e^2}{\hbar^2}\int[d\bm k] \sum_n \Bigg[-\frac{\beta}{2\cosh^2(\frac{\beta E_{n}}{2})} \nonumber
\\
+ &\frac{\tanh(\frac{\beta E_{n}}{2})}{E_{n}} \Bigg] \frac{\Delta^2}{E_{n}^2} \partial_a \varepsilon_{n} \partial_b \varepsilon_{n}, \label{D_conv}
\\
D_{\text{geom}}^{ab}&=\frac{e^2\Delta^2}{\hbar^2}\int[d\bm k] \sum_{n\neq m} \Bigg[ \frac{\tanh(\frac{\beta E_{n}}{2})}{E_{n}} \nonumber
\\
&- \frac{\tanh(\frac{\beta E_{m}}{2})}{E_{m}} \Bigg] \frac{(\varepsilon_{n}-\varepsilon_{m})^2}{E_{m}^2-E_{n}^2} \label{D_geom}
\\
&\times \text{Re}\,\mathcal{A}_{nm}^a\mathcal{A}_{mn}^b, \nonumber
\end{align}
where $\varepsilon_{n}$ is the single-electron spectrum corresponding to the wave function $|n\rangle$, $E_{n}=\sqrt{(\varepsilon_{n}-\mu)^2+\Delta^2}$ is the quasiparticle spectrum; and $\beta=1/k_BT$. We note that $D_{\text{conv}}^{ab}$ is associated with the diagonal part of the current operator and is the combination of the superfluid weight of each single band [Eq.~(\ref{D_sing})], while the $D_{\text{geom}}^{ab}$ relies on the off-diagonal part of the current operator and exhibits a geometric origin \cite{peotta2015, lianglong2017, huhtinen2022}. Remarkably, in the isolated band limit, where the partially filled flat band $\varepsilon_{n^*}$ is well separated from the other bands $\varepsilon_{m}$ ($m\neq n^*$) by a gap much larger than $\Delta$, the conventional term [Eq.~(\ref{D_conv})] vanishes while the geometric term is greatly simplified to \cite{peotta2015, xiefang2020}
\begin{equation} \label{D_geom_iso}
D_\text{geom}^{ab}=\frac{4e^2\Delta\sqrt{\nu(1-\nu)}}{\hbar^2} \int[d\bm k] g_n^{ab},
\end{equation}
\textcolor{black}{where $\nu$ is the filling ratio of the isolated flat band.} Similar to the effective mass [Eq.~(\ref{m_eff})], the superfluid weight [Eq.~(\ref{D_geom_iso})] is also determined by a Brillouin zone integral of the quantum metric. This unambiguously confirms the presence of a finite-sized supercurrent.

\subsection{\textcolor{black}{Application to twisted bilayer graphene}}
\begin{figure*}[!t]
\centering
\includegraphics[width=140mm]{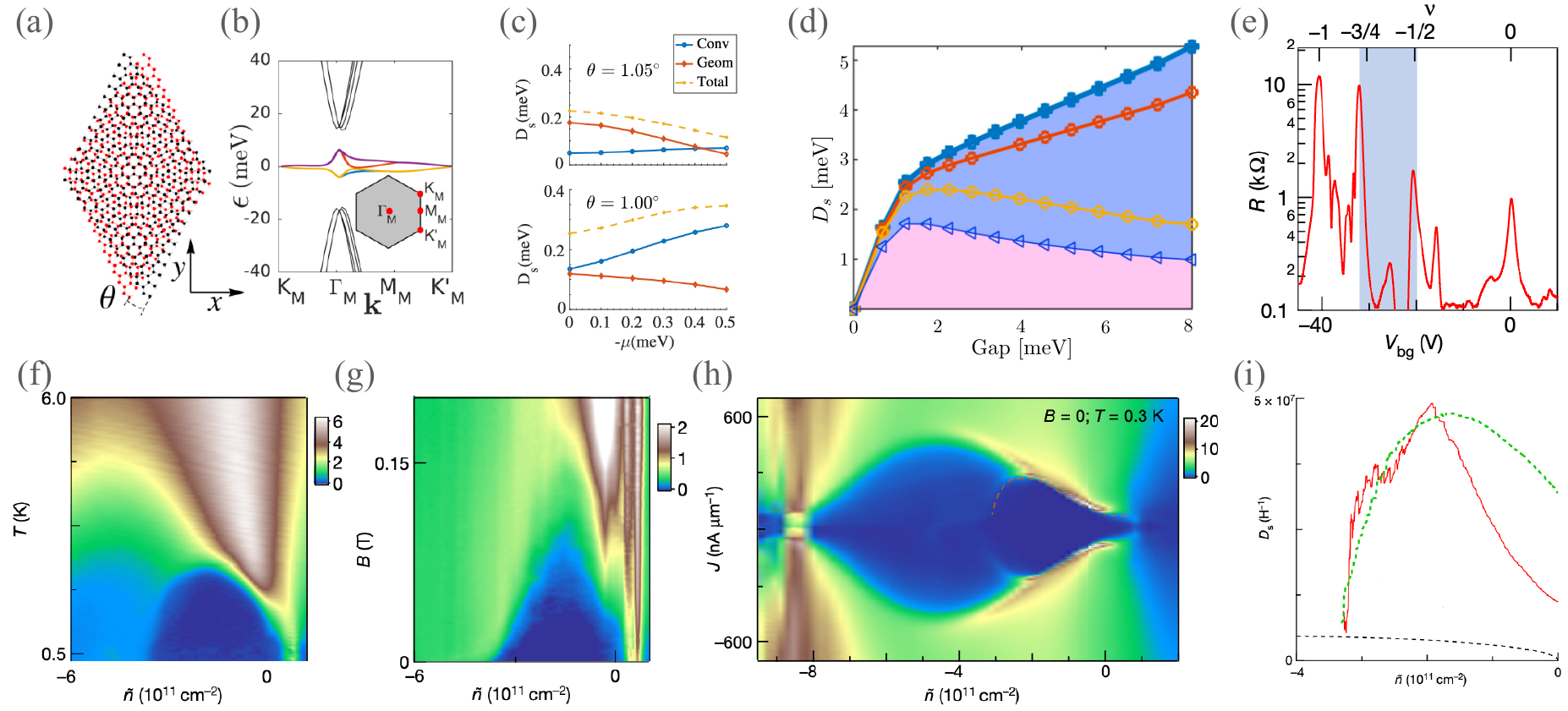}
\caption{(a) Lattice structure of twisted bilayer graphene. (b) Band structure of twisted bilayer graphene plotted along the high-symmetry path. Inset: the \textcolor{black}{Moir\'e} Brillouin zone. The nearly flat bands are highlighted with colors. (c) Superfluid weight versus filling at the magic angle $\theta=1.05^\circ$ and $\theta=1.00^\circ$. (d) Superfluid weight versus superconductor gap. Thick blue: full superfluid weight. Red: superfluid weight calculated with 8 lowest energy bands (4 dispersive, 4 nearly flat). Yellow: superfluid weight calculated with 4 nearly flat bands. Blue: Conventional superfluid weight. (e) Longitudinal resistance versus back gate voltage of twisted bilayer graphene. Shade marks the filling where superconductivity occurs. (f) Resistivity on the $\tilde n -T$ plane shows a superconducting dome (dark blue). (g) Resistivity on the $\tilde n -B$ plane shows a superconducting dome (dark blue). (h) Differential resistivity on the $\tilde n -J$ plane shows a superconducting dome (dark blue). The dome boundary (red dashed curve) marks the critical supercurrent. (i) Superfluid weight calculated with critical supercurrent (red) through Eq.~(\ref{D_exp}) is well fitted by including the geometric contribution $D=0.33\tfrac{(2\pi e)^2}{h^2}\Delta$ (green) and dominates over the conventional contribution approximated by $D=\tilde n e^2/m^*$ (black). Panels (a, b, d) are adapted from Ref.~\cite{julku2020}. Panel (c) is adapted from Ref.~\cite{huxiang2019}. Panels (e-i) are adapted from Ref.~\cite{tianhaidong2023}.
}
\label{fig6}
\end{figure*}

One prominent example of flat-band superconductors is the twisted bilayer graphene \cite{caoyuan2018a} [Fig.~\ref{fig6}(a)], which exhibits nearly flat \textcolor{black}{Moir\'e} bands at the magic angle $\sim1.05^\circ$ [Fig.~\ref{fig6}(b)]. The superfluid weight of twisted bilayer graphene is bounded from below by the $C_{2z}T$ Wilson loop winding number arising from the nontrivial topology of the two lowest flat bands \cite{xiefang2020}. In the presence of nearest-neighbor pairing mechanism [cf., the onsite pairing in Eq.~(\ref{hubbard})], nematic superconductivity emerges and anisotropy is found in the superfluid weight \cite{julku2020}. Since the flat bands of twisted bilayer graphene is not perfectly dispersionless but have a band width of several milli-electronvolts [Fig.~\ref{fig6}(b)], the conventional contribution $D_\text{conv}^{ab}$ [Eq.~(\ref{D_conv})] cannot be completely suppressed and may surpass the geometric contribution  $D_\text{geom}^{ab}$ [Eq.~(\ref{D_geom})] at certain fillings [Fig.~\ref{fig6}(c)]. $D_\text{conv}^{ab}$ can even dominate the $D_\text{geom}^{ab}$ when the twist angle is slightly tuned away from the magic angle [Fig.~\ref{fig6}(c)], suggesting that the geometric effects in twist bilayer graphene sensitively depends on the flatness of the bands \cite{huxiang2019}. Moreover, the geometric superfluid weight of twisted bilayer graphene grows linearly with respect to the gap [Fig.~\ref{fig6}(d)], consistent with the prediction of Eq.~(\ref{D_geom_iso}).

For a twisted bilayer graphene device with a twist angle $\theta=1.08^\circ\pm0.02^\circ$ \cite{tianhaidong2023}, its longitudinal resistance versus back gate voltage exhibits peaks at $\nu=0$, $-1/2$, $-3/4$, and $-1$ [Fig.~\ref{fig6}(e)]. The Dirac revival is found at $\nu=-1/2$, where the effective charge density $\tilde n=0$ \cite{tianhaidong2023}. For $-3.5<\tilde n<0.3\times 10^{11}\,\text{cm}^{-2}$, which roughly corresponds to the filling regime $-5/8<\nu<-1/2$, a superconducting dome is observed in the $\tilde n-T$ plane [Fig.~\ref{fig6}(f)]. The dome is approximately peaked at the optimal doping $\tilde n_\text{op}=-1.8\times 10^{11}\,\text{cm}^{-2}$ with a Berezinskii-Kosterlitz-Thouless transition temperature $T_c=2.2\,\text K$. In the $\tilde n-B$ plane, a similar dome is observed, revealing an upper critical field $B_{c2}=0.1\,\text T$ at the optimal doping [Fig.~\ref{fig6}(g)]. The superconducting coherence length $\xi=\sqrt{\Phi_0/2\pi B_{c2}}$ is calculated to be approximately $57\,\text{nm}$ at $\tilde n_\text{op}$. This suggests at the optimal doping a superconductor gap \textcolor{black}{$\Delta=\hbar v_F/\pi\xi\approx 0.0037\,\text{meV}$}, where $v_F\approx 10^3\,\text{m/s}$ is adopted \cite{tianhaidong2023}. Consequently, the ratio \textcolor{black}{$\Delta/k_BT_c\approx 0.02$} is reached, far from the BCS value 1.764 \cite{tinkham2004}. Alternatively, estimating the effective mass at the optimal doping according to $m^*=\hbar k_F/v_F$ ($k_F=\sqrt{2\pi\tilde n_\text{op}}$ being the Fermi momentum), the superfluid weight \textcolor{black}{reads $D=\tilde n_\text{op}e^2/m^* \approx 4.12\times10^6\,\text{H}^{-1}$.} The Nelson-Kosterlitz criterion \cite{xiefang2020} then gives an estimate of the transition temperature
\begin{equation} \label{Tc}
T_c\leq\frac{\pi\hbar^2}{8e^2k_B}D\approx 0.05\,\text{K},
\end{equation}
which is far less than the measured $T_c=2.2\,\text{K}$. Since Eq.~(\ref{Tc}) underestimates the superfluid weight by neglecting the geometric contribution [Eq.~(\ref{D_geom})], its pronounced deviation from the measured $T_c=2.2\,\text{K}$ justifies that the quantum geometric effect greatly contributes to the superconductivity of twisted bilayer graphene.

To scrutinize the quantum geometric effect on the superfluid weight, it is instructive to examine the differential resistivity on the $\tilde n-J$ plane, where a superconducting dome is observed [Fig.~\ref{fig6}(h)]. The critical current [i.e., boundary of the dome, see red dashed curve in Fig.~\ref{fig6}(h)] yields a superfluid weight \cite{tianhaidong2023}
\begin{equation} \label{D_exp}
D=\frac{2\pi\xi}{\Phi_0}J_{cs}.
\end{equation}
which also exhibits a dome structure. The dome top is located at $5\times10^7\,\text{H}^{-1}$ [red curve, Fig.~\ref{fig6}(i)], far greater than the estimated conventional contribution [black curve, Fig.~\ref{fig6}(i)], and should be dominantly contributed by the geometric effect [Eq.~(\ref{D_geom_iso})], which is parameterized as $D_\text{geom}=b\tfrac{e^2}{\hbar^2}\Delta$. The $D$ versus $\tilde n$ curve at $b=0.33$ [green curve, Fig.~\ref{fig6}(i)] well fits the experimental data [red curve, Fig.~\ref{fig6}(i)] and in general captures the overall trend. Such a fit straightforwardly confirms the crucial role of quantum geometry in the superconductivity of twisted bilayer graphene.

\section{FRACTIONAL CHERN INSULATORS}
The quantization of Hall conductivity of two-dimensional electron gas, $\sigma=\mathcal C\frac{e^2}{\hbar}$, is a prominent example violating the Landau phase transition theory \cite{klitzing1980}. This integer quantum Hall effect is characterized by the topological invariant $\mathcal C$, which is known as the Chern number and is originated from the Landau levels produced by the applied magnetic field \cite{thouless1982}. Intriguingly, the Chern number may also arise from the magnetization or spin-orbit coupling of materials in the complete absence of Landau levels or magnetic fields \cite{haldane1988}. Such materials, referred to as the Chern insulators, also exhibit quantized Hall conductance, because they share exactly the same \textcolor{black}{topological} feature as the integer quantum Hall effect.

Strong electron-electron interaction can lift the degeneracy of the Landau levels in the integer quantum Hall effect and the partial filling of such Landau levels produces a fractional quantum Hall effect \cite{tsui1982, laughlin1983}. The observation of the fractionally quantized Hall conductance in two-dimensional electron gas requires an extremely strong magnetic field, typically on the order of several tens of tesla \cite{tsui1982}, and thus causes great difficulty in the experimental implementation. Hence, a natural question arises: can the fractionally quantized Hall conductance be visualized in zero magnetic field? Or, equivalently, by referring to the relationship between the integer quantum Hall effect and the Chern insulator, can one construct from the fractional quantum Hall effect a fractional Chern insulator?

Following the construction of Chern insulators \cite{haldane1988}, the necessary ingredients of a fractional Chern insulator seem to be a nearly flat Chern band, which encodes the strong electron-electron interaction and possesses the required topology. It is shown that these two requirements can be simultaneously satisfied in tight-binding models with proper short-ranged hoppings \cite{tang2011, sunkai2011, neupert2011}. Such fractional Chern insulators host nearly flat Bloch bands with $\mathcal C=\pm 1$, thus highly mimicking the Landau levels in the fractional quantum Hall effect. However, the experimental implementation of such fractional Chern insulator is challengeable. It is then realized that the challenge is deeply rooted in the key difference between flat Chern bands and Landau levels: the quantum geometry of flat Chern bands (Landau levels) is in general fluctuating (homogeneous) in the Brillouin zone; and the stability of the fractional Chern insulating phase (i.e., the fractional quantum anomalous Hall effect) sensitively depends on the how uniformly the quantum geometry is distributed \textcolor{black}{in the Brillouin zone} \cite{parameswaran2012, roy2014, jackson2015}. 

\textcolor{black}{Besides the requirement on momentum-space distribution, it is worth noting that the quantum metric tensor and Berry curvature for a given Bloch band (labelled with $n$) are governed by Eq. (\ref{Eq: inequality}). For the lowest Landau level in the fractional quantum Hall effect, both inequalities adopt the equal signs \cite{wangjie2021}. We thus expect the flat Chern band mimicking the lowest Landau level to act in the same way  or at least approximately satisfies $\text{Tr}\,\bm {g}_n \simeq |\boldsymbol{\Omega}_n |$ and $\text{Det}\, \bm {g}_n  \simeq |\boldsymbol{\Omega}_n |^2/4$.}

To sum up, a stabilized fractional Chern insulator follows these three criteria: (i) A flat or nearly flat band, whose kinetic energy is overwhelmed by the electron-electron interaction; (ii) A nontrivial Chern number which is either intrinsic or interaction-induced; and (iii) \textcolor{black}{Quantum geometry that is almost uniformly distributed in the Brillouin zone} \cite{parameswaran2012, roy2014, jackson2015} \textcolor{black}{and approximately satisfies the trace and determinant conditions if mimicking the lowest Landau level}. The simultaneous satisfaction of all the three criteria is highly nontrivial and in general requires platforms with great tunability. Fortunately, the \textcolor{black}{Moir\'e} materials shed new light on the accessibility of such platforms \cite{ledwith2020, repellin2020, abouelkomsan2020, wilhelm2021, xieyonglong2021, spanton2018, luzhengguang2024, caijiaqi2023, zengyihang2023, park2023, xufan2023}.

\subsection{Magic-angle twisted bilayer graphene}
The magic-angle twisted bilayer graphene has been known to exhibit nearly flat bands and thus can support strong electron correlation effects \cite{bistritzer2011}. The correlated Chern insulating phase further confirms the required nontrivial topology for the fractional Chern insulator \cite{serlin2020, sharpe2019, nuckolls2020, choi2021, park2021a}. In the chiral limit (i.e., the same sublattices are decoupled across the two honeycomb layers), the flat bands of magic-angle twisted bilayer graphene acquire ideal quantum geometry \cite{ledwith2020}
\begin{equation}
g_n^{ab} =\frac 1 2 \delta_{ab} |\boldsymbol{\Omega }_n|,
\end{equation}
which means that both inequalities in Eq.~(\ref{Eq: inequality}) adopt the equal signs, same as the lowest Landau level in the fractional quantum Hall effect. Considering that the realistic magic-angle twisted bilayer graphene does not qualitatively differ from its chiral limit \cite{tarnopolsky2019}, one naturally expects the magic-angle twisted bilayer graphene to have nearly ideal quantum geometry, thus in principle satisfying all the three aforementioned criteria for a fractional Chern insulator.

\begin{figure*}[!t]
\centering
\includegraphics[width=140mm]{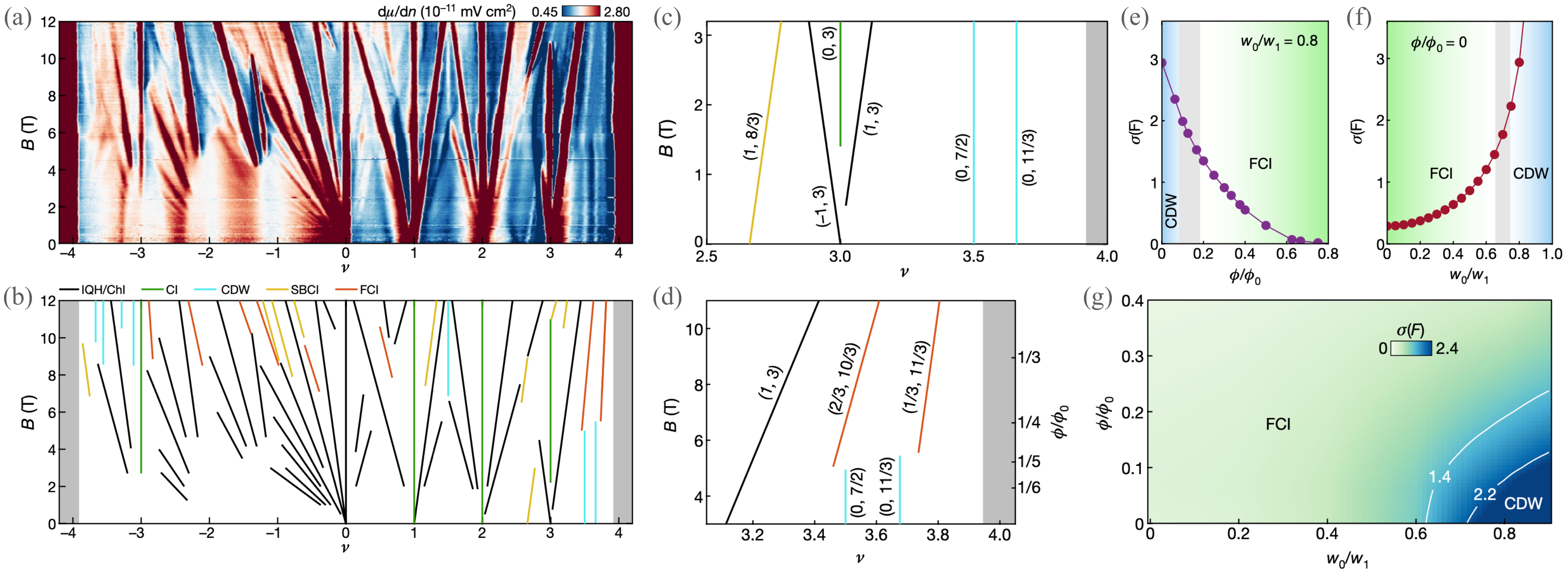}
\caption{(a) Local inverse compressibility $d\mu/dn$ on the $\nu-B$ plane. $\nu$ is the filling per \textcolor{black}{Moir\'e} unit cell and $B$ is the applied magnetic field. (b) The line trajectories in the inverse compressibility measurement are classified as Chern insulators/integer quantum Hall states (black), trivial correlated insulators (green), charge density waves (cyan), translational symmetry-broken Chern insulators (yellow), and fractional Chern insulators (red). (c) Zoom-in of panel (b) in the low-field regime. (d) Zoom-in of panel (b) in the high-field regime. (e) Standard deviation of Berry curvature as a function of magnetic field for a realistic magic-angle twist bilayer graphene with $w_0/w_1=0.8$. Fractional Chern insulating (CDW) phase is observed at the high-field (low-field) regime. The onset of the fractional Chern insulating phase is approximately located at $(\phi/\phi_0)_c\simeq 0.18$ and $\sigma_c(F)=1.4$. (f) Standard deviation of Berry curvature as a function of the ratio $w_0/w_1$ of twisted bilayer graphene at zero applied magnetic field. The onset of the fractional Chern insulating phase is approximately at the ratio $w_0/w_1=0.65$. (g) The phase diagram of twisted bilayer graphene on the $w_0/w_1-\phi/\phi_0$ plane. CDW is located at the lower right corner. The transition to fractional Chern insulators occurs in the regime $1.4\leq \sigma(F) \leq 2.2$. All the panels are adapted from Ref.~\cite{xieyonglong2021}.
}
\label{fig7}
\end{figure*}

It has been predicted that fractional Chern insulating states occur in devices comprising of magic-angle twisted bilayer graphene aligned with hexagonal boron nitride \cite{ledwith2020, repellin2020, abouelkomsan2020, wilhelm2021}. Performing local electronic compressibility measurements on such a device (with a twist angle $\sim1.06^\circ$) using a scanning single electron transistor \cite{xieyonglong2021}, the inverse compressibility $d\mu/dn$ can be obtained and plotted on the $\nu-B$ plane [Fig.~\ref{fig7}(a)], where $\nu$ is the \textcolor{black}{Moir\'e} band filling factor and $B$ is the applied magnetic field. The applied magnetic field is used to properly tailor the distribution of quantum geometry, as will be analyzed later. A variety of incompressible states are visualized as linear trajectories in Figs.~\ref{fig7}(a, b). These trajectories follow the Diophantine equation \cite{macdonald1983}
\begin{equation} \label{diophantine}
\nu=t\frac{\phi}{\phi_0}+s,
\end{equation}
where $\phi$ is the magnetic flux per \textcolor{black}{Moir\'e} unit cell, $\phi_0$ is the flux quantum, and $(t,s)$ are a pair of parameters. The values of $(t,s)$ determine the nature of the incompressible states [Chern insulators/integer quantum Hall states, correlated insulators, charge density waves (CDWs), translational symmetry broken Chern insulators, fractional Chern insulators, etc.], as detailed in Table~\ref{tab1}.

\begin{table}[t]%
\vspace*{-2pt}
\centering
\tablefont
\caption{Classification of incompressible states in magic-angle twisted bilayer graphene aligned with hexagonal boron nitride \cite{xieyonglong2021}.\label{tab1}}%
\begin{tabular*}{\columnwidth}{@{\extracolsep\fill}lcc@{\extracolsep\fill}}%
\toprule
{Incompressible states} & {Values of $(t,s)$}  \\
\colrule
Integer quantum Hall states & Integer $t\neq0$, $s=0$ 
\\
Chern insulators & Integer $t\neq0$, integer $s\neq 0$   
\\
Correlated insulators & $t=0$, integer $s\neq 0$ 
\\
Charge density waves & $t=0$, fractional $s$ 
\\
TS-broken Chern insulators & Integer $t\neq0$, fractional $s$ 
\\
Fractional Chern insualtors & fractional $t$, fractional $s$
\\
\botrule
\end{tabular*}\vspace*{-3pt}
\end{table}

To scrutinize the incompressible states, we first examine the low-field regime near $\nu=3$. Two Chern insulating states $(\pm 1, 3)$ and a correlated insulating state $(0, 3)$ are emanating from $\nu=3$ at zero magnetic field [Fig.~\ref{fig7}(c)]. A symmetry-broken Chern insulating state is observed on the left with $(1, 8/3)$ [Fig.~\ref{fig7}(c)]. Besides the insulating states, two CDW states with $(0, 7/2)$ and $(0, 11/3)$ are found. Remarkably, these two CDW states terminate around $B\approx5\,\text T$, which is exactly the rise of the two fractional Chern insulating states with $(2/3, 10/3)$ and $(1/3, 11/3)$  [Fig.~\ref{fig7}(d)], implying the competition between the CDW and fractional Chern insulating states \cite{wilhelm2021, kourtis2012}. This competition arises from the fact that fractional Chern insulators are stabilized by a uniform distribution of quantum geometry  while CDW is facilitated by fluctuating quantum geometry \cite{wilhelm2021}, e.g., strongly localized Berry curvature at the center of the Brillouin zone \cite{pierce2021}. Therefore, it is instructive to study the transition from CDW to fractional Chern insulators by checking the \textcolor{black}{momentum-space} distribution homogeneity of Berry curvature, characterized by its standard deviation $\sigma(F)$.

To figure out the critical value of $\sigma(F)$, we note that the interplay between the CDW and fractional Chern insulators in twisted bilayer graphene has already been found to rely on the ratio $w_0/w_1$, where $w_0$ ($w_1$) is the interlayer tunneling between identical (different) sublattices of the twisted layers \cite{abouelkomsan2020, wilhelm2021, repellin2020}.  For twisted bilayer graphene, $w_0/w_1\simeq0.8$ \cite{koshino2018} and $\sigma(F)$ decreases with the increased magnetic field [Fig.~\ref{fig7}(e)]. This reveals that the effect of magnetic field here is to tailor the quantum geometry, making it more uniformly distributed \textcolor{black}{in the Brillouin zone}, as mentioned above. As illustrated in Fig.~\ref{fig7}(d), the onset of fractional Chern insulator is $B\simeq 5\,\text T$, i.e., $(\phi/\phi_0)_c\simeq 0.18$, the critical value of $\sigma(F)$ can thus be read off from Fig.~\ref{fig7}(e) as $\sigma_c(F)\simeq1.4$, which marks the transition from fractional Chern insulators to CDW phase. In the absence of magnetic fields, $\sigma(F)$ has a positive dependence on the ratio $w_0/w_1$ [Fig.~\ref{fig7}(f)]. The transition $\sigma_c(F)\simeq1.4$ is approximately located at $(w_0/w_1)_c=0.65$, implying that a chiral (normal) magic-angle twisted bilayer graphene, labelled by $w_0/w_1=0$ ($w_0/w_1=0.8$), is in the fractional Chern insulating (CDW) phase. On a more generic view, a phase diagram of twisted bilayer graphene can be constructed on the $w_0/w_1-\phi/\phi_0$ plane, where the CDW phase is located near the lower right corner with large $w_0/w_1$ values but small $\phi/\phi_0$ values [Fig.~\ref{fig7}(g)]. Field-free magic-angle twisted bilayer graphene belongs to this regime, but can be fortunately tuned to the adjacent fractional Chern insulating phase with a moderate magnetic field $B\simeq 5\,\text T$.

\subsection{Other \textcolor{black}{Moir\'e} platforms}
Prior to magic-angle twisted bilayer graphene aligned with hexagonal boron nitride, fractional Chern insulating states are proposed, for the first instance, in a dual-gated Bernal bilayer graphene device \cite{spanton2018}, where a rotational alignment $\sim1^\circ$ is prepared between the Bernal bilayer graphene and one of the gate. The measurement of penetration field capacitance distinguishes the incompressible and compressible states and reveals the fractional Chern insulating states, which arise from the Harper-Hofstadter bands produced by the interplay between the \textcolor{black}{Moir\'e} superlattice potential and the applied magnetic field \cite{anderson2023}. However, the required magnetic field is quite strong $B\sim 30\,\text T$, because its effect here is creating appropriate topology rather than slightly tailoring the quantum geometry as in the magic-angle twisted bilayer graphene \cite{xieyonglong2021}.

\begin{figure*}[!t]
\centering
\includegraphics[width=140mm]{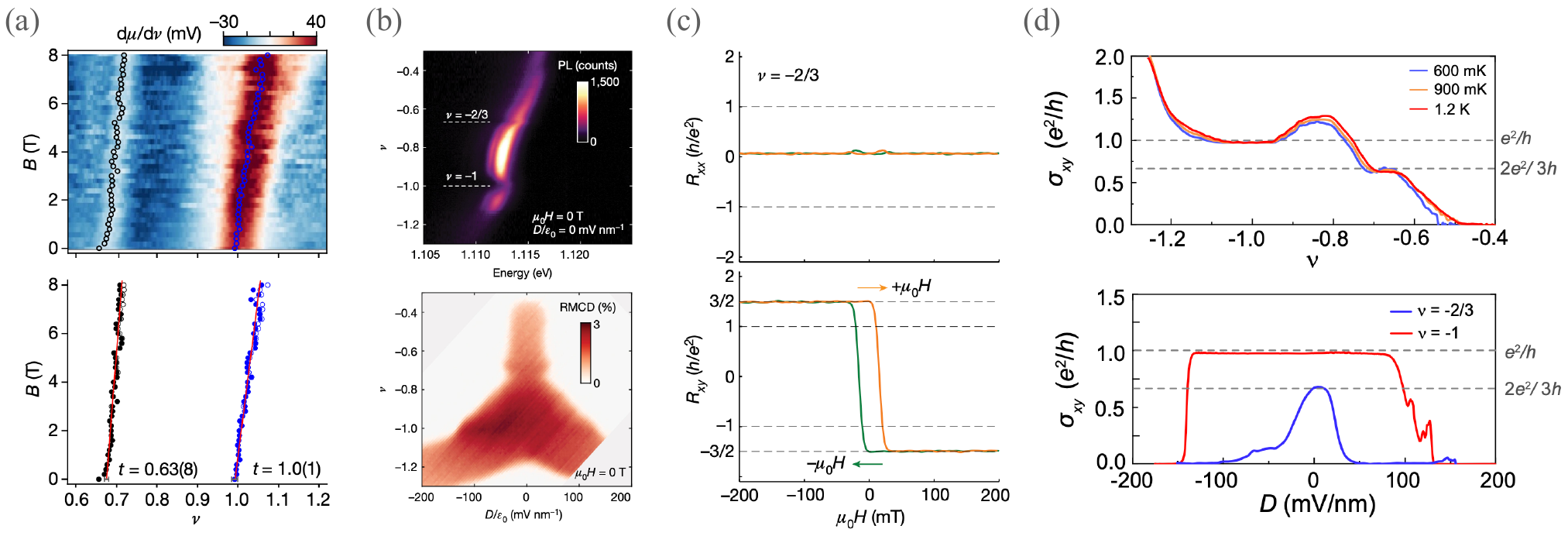}
\caption{(a) Incompressibility measurement of twisted bilayer MoTe$_2$. Integer (blue) and fractional (black) Chern insulators are observed respectively for hole filling $\nu=1$ and $\nu=2/3$. Adapted from Ref.~\cite{zengyihang2023}. (b) Trion photoluminescence exhibits blueshifts at $\nu=-1$ and $\nu=-2/3$, respectively indicating integer and fractional Chern insulating states, which are also observable in the reflective magnetic circular dichroism measurement. Adapted from Ref.~\cite{caijiaqi2023}. (c) Longitudinal and Hall resistances versus magnetic field in twisted bilayer MoTe$_2$. Plateaus are observed at $R_{xy}=\pm{3h}/{2e^2}$, corresponding to fractional filling $\nu=-2/3$. Adapted from Ref.~\cite{park2023}. (d) Hall conductance versus filling/displacement in twisted bilayer MoTe$_2$. Fractional Hall conductance $\sigma_{xy}={2e^2}/{3h}$ is the smoking-gun evidence of fractional Chern insulators. Adapted from Ref.~\cite{xufan2023}.  }
\label{fig8}
\end{figure*}

The bilayer graphene-based fractional Chern insulators require magnetic fields to tune the topology or the quantum geometry \cite{spanton2018, xieyonglong2021}, thus cannot support the fractional quantum anomalous Hall effect. By contrast, \textcolor{black}{Moir\'e} materials based on transition metal dichalcogenides have been predicted to simultaneously host proper interaction, topology, and quantum geometry \cite{devakul2021, crepel2023, liheqiu2021, morales2023} and thus may be potential candidates of fractional Chern insulators at zero magnetic field. In particular, twisted bilayer MoTe$_2$ has been confirmed to exhibit the fractional quantum anomalous Hall effect \cite{zengyihang2023, caijiaqi2023, park2023, xufan2023}. The incompressibility measurement of twisted bilayer MoTe$_2$ has revealed both integer and fractional quantum anomalous Hall states respectively at $1$ and $2/3$ hole filling \cite{zengyihang2023} [Fig.~\ref{fig8}(a)], while the same states are also visualized with trion photoluminescence and reflective magnetic circular dichroism measurements \cite{caijiaqi2023} [Fig.~\ref{fig8}(b)]. Moreover, the direct transport measurements have unveiled fractional Hall resistance \cite{park2023} [Fig.~\ref{fig8}(c)] and conductance \cite{xufan2023} [Fig.~\ref{fig8}(d)], providing smoking-gun evidences of fractional Chern insulators at zero fields.

\section{CONCLUSIONS}
We present a comprehensive review of the quantum geometric effect on the transport, superconductivity, and topology of condensed matter. These three subjects are also nicely reviewed in Refs.~\cite{maqiong2021}, Refs.~\cite{peotta2023, torma2022}, and Refs.~\cite{bergholtz2013, liuzhao2024}, respectively.

First, we review that quantum geometry yields a variety of types of nonlinear transport, such as the electric nonlinear Hall effect \cite{sodemann2015, duzz2018, duzz2019,  maqiong2019, kangkaifei2019, huangmeizhen2023, tiwari2021, mateng2022, shvetsov2019, kumar2021, dzsaber2021, gaoyang2014, kaplan2024, wangchong2021, liuhuiying2021, gaoanyuan2023, wangnaizhou2023}, magneto-nonlinear Hall effect \cite{gaoyang2014, wanglujunyu2024}, and nonlinear longitudinal conductivity \cite{kaplan2024, wangnaizhou2023}. The electric nonlinear Hall effect may arise from the dipole moment of either Berry curvature \cite{sodemann2015, duzz2018, duzz2019, maqiong2019, kangkaifei2019, huangmeizhen2023, tiwari2021, mateng2022, shvetsov2019, kumar2021, dzsaber2021} or quantum metric \cite{gaoyang2014, kaplan2024, wangchong2021, liuhuiying2021,  gaoanyuan2023, wangnaizhou2023}, depending on the symmetry. The magneto-nonlinear Hall effect is contributed by the spin/orbital magnetic moment and anomalous spin/orbital polarizability \cite{gaoyang2014, wanghui2024, wanglujunyu2024}. The longitudinal nonlinear conductivity relies on the quantum metric and thus may co-exist with the electric nonlinear Hall effect \cite{kaplan2024, wangnaizhou2023}.

Second, we examine the quantum geometry in flat-band superconductors, paying close attention to the superfluid weight \cite{peotta2015, julku2016, lianglong2017, torma2018, huhtinen2022}. Even though the electrons on the flat bands are inert, transport is still possible and arises from the quantum metric of the bands. The measured critical supercurrent greatly surpasses that estimated in the BCS formalism \cite{tianhaidong2023}. The difference results from the interband quantum metric, which gives rise to a geometric superfluid weight \cite{peotta2015, xiefang2020}. The geometric superfluid weight can even survive when the bands become completely flat and is responsible for the enhanced transition temperature of flat-band superconductors.

Lastly, we summarize the important role of quantum geometry in the fractional Chern insulators. Besides the interacting Chern bands, fractional Chern insulators also require almost uniformly distributed \textcolor{black}{momentum-space} quantum geometry, which is responsible for the stabilization of the fractional Chern insulating phase \cite{parameswaran2012, roy2014, jackson2015}. The simultaneous constraints on interaction, topology, and quantum geometry indicate that fractional Chern insulators may only be realized in highly tunable platforms. The rise of \textcolor{black}{Moir\'e} materials has presented several candidates of fractional Chern insulators \cite{ledwith2020, repellin2020, abouelkomsan2020, wilhelm2021, xieyonglong2021, spanton2018, luzhengguang2024, caijiaqi2023, zengyihang2023, park2023, xufan2023}. While Bernal bilayer graphene-hexagonal boron nitride \textcolor{black}{Moir\'e} heterostructure \cite{spanton2018} and magic-angle twisted bilayer graphene \cite{xieyonglong2021} require magnetic field to respectively tailor the topology and quantum geometry, rhombohedral pentalayer graphene–hexagonal boron nitride \textcolor{black}{Moir\'e} devices \cite{luzhengguang2024} and twisted bilayer MoTe$_2$ \cite{caijiaqi2023, zengyihang2023, park2023, xufan2023} do exhibit fractional quantum anomalous Hall effects.

\section{OUTLOOK}
The area of quantum geometry is rapidly expanding, leaving many new directions to be further studied. 

In the nonlinear transport, \textcolor{black}{disorder has been known to interplay with quantum geometry \cite{duzz2019, atencia2022, atencia2023}. In particular, Berry curvature can enter certain parts of the electric nonlinear Hall conductivities arising from side jump and skew scattering \cite{duzz2019}.  It will thus be interesting to check how the disorder interacts with quantum metric and affects the electric nonlinear Hall and longitudinal conductivities. The interplay between disorder and the anomalous spin/orbital polarizability should lead to a comprehensive understanding of the magneto-nonlinear Hall effect and thus deserves a careful study.} \textcolor{black}{Moreover, it may be instructive to examine the role of quantum geometry in nonlinear transport beyond the electric and magneto- regimes (e.g., spin regime \cite{liuhong2023, bhalla2021}) and} in higher-order nonlinear transport \cite{laishen2021, liuhuiying2022}, in which new types of quantum geometric quantities may be engaged.

In flat-band superconductors such as magic-angle twisted bilayer graphene, the evidence of quantum geometric effect has been confirmed in a qualitative fashion. A more quantitative check is preferred. One could also investigate the quantum geometry in other \textcolor{black}{Moir\'e} superconductors such as twisted trilayer graphene \cite{park2021b} and twisted double bilayer graphene \cite{suruiheng2023}. Additionally, \textcolor{black}{Moir\'e} superconductors are playgrounds of novel pairing symmetries (e.g., $d+id$ pairing in twisted cuprates \cite{can2021, zhaofrank2023}). Studying how such pairing symmetries interplay with quantum geometry could be crucial for understanding the nature of the hosting superconductors.

In the context of fractional Chern insulators, one of the most important applications is topological quantum computation. For such a purpose, a great variety of platforms, especially \textcolor{black}{Moir\'e} materials have to be explored to identify a suitable non-Abelian candidate \cite{liuzhao2019}. Such non-Abelian fractional Chern insulators should have appropriate quantum geometry requiring no further tuning by magnetic fields (e.g., \textcolor{black}{twisted} MoTe$_2$ \cite{caijiaqi2023, zengyihang2023, park2023, xufan2023} and pentalayer graphene \cite{luzhengguang2024}) but preferably operate at a higher temperature. \textcolor{black}{Furthermore, Moir\'e materials like twisted MoTe$_2$ can host additional many-body states, exemplified by the anomalous composite Fermi liquid \cite{goldman2023, dong2023}. The interplay between such many-body states and the fractional Chern insulating states had better be settled. And a quantum geometric theory unifying all these states is highly preferred.
}

\section{ACKNOWLEDGEMENTS}
We thank Qian Niu, Yang Gao, and Shengyuan Yang for the insightful discussions. 

\section{FUNDING}
This work was supported by the National Key R\&D Program of China (2022YFA1403700), Innovation Program for Quantum Science and Technology (2021ZD0302400), the National Natural Science Foundation of China (12304196, 12350402, and 11925402), Guangdong Basic and Applied Basic Research Foundation (2022A1515111034, 2023B0303000011), Guangdong province (2020KCXTD001), the Science, Technology and Innovation Commission of Shenzhen Municipality (ZDSYS20190902092905285), and Center for Computational Science and Engineering of SUSTech.

\section{AUTHOR CONTRIBUTIONS}
H.-Z. L. conceived the project. H.-Z. L. and X. C. X. oversaw all research. T. L. reviewed quantum geometry in nonlinear transport, flat-band superconductors, and fractional Chern insulators. X.-B. Q. reviewed the physical and mathematical backgrounds of quantum geometry. T. L. and X.-B. Q. wrote the manuscript under supervision of H.-Z. L. and X. C. X.. All the authors participated in analyzing literatures and revising the manuscript.
\\[1em]
\noindent\textbf{\textit{Conflict of interest statement.}} None declared.

\bibliographystyle{nsr}
\bibliography{quantum_geometry}

\end{document}